# Probabilities of Collisions of Planetesimals from Different Regions of the Feeding Zone of the Terrestrial Planets with the Forming Planets and the Moon

## S. I. Ipatov*

*Vernadsky Institute of Geochemistry and Analytical Chemistry, Russian Academy of Sciences, Moscow, 119991 Russia *e-mail: siipatov@hotmail.com*

**Abstract**—Migration of planetesimals from the feeding zone of the terrestrial planets, which was divided into seven regions depending on the distance to the Sun, was simulated. The influence of gravity of all planets was taken into account. In some cases, the embryos of the terrestrial planets rather than the planets themselves were considered; their masses were assumed to be 0.1 or 0.3 of the current masses of the planets. The arrays of orbital elements of migrated planetesimals were used to calculate the probabilities of their collisions with the planets, the Moon, or their embryos. As distinct from the earlier modeling of the evolution of disks of the bodies coagulating in collisions, this approach makes it possible to calculate more accurately the probabilities of collisions of planetesimals with planetary embryos of different masses for some evolution stages. When studying the composition of planetary embryos formed from planetesimals, which initially were at different distances from the Sun, we considered the narrower zones, from which planetesimals came, as compared to those examined earlier, and analyzed the temporal changes in the composition of planetary embryos rather than only the final composition of planets. Based on our calculations, we drew conclusions on the process of accumulation of the terrestrial planets. The embryos of the terrestrial planets, the masses of which did not exceed a tenth of the current planetary masses, accumulated planetesimals mainly from the vicinity of their orbits. When planetesimals fell onto the embryos of the terrestrial planets from the feeding zone of Jupiter and Saturn, these embryos had not yet acquired the current masses of the planets, and the material of this zone (including water and volatiles) could be accumulated in the inner layers of the terrestrial planets and the Moon. For planetesimals which initially were at a distance of 0.7–0.9 AU from the Sun, the probabilities of their infall onto the embryos of the Earth and Venus, the mass of which is 0.3 of the present masses of the planets, differed less than twofold for these embryos. The total mass of planetesimals, which initially were in each part of the region between 0.7 and 1.5 AU from the Sun and collided with the almost-formed Earth and Venus, apparently differed by less than two times for these planets. The inner layers of each of the terrestrial planets were mainly formed from the material located in the vicinity of the orbit of a certain planet. The outer layers of the Earth and Venus could accumulate the same material for these two planets from different parts of the feeding zone of the terrestrial planets. The Earth and Venus could acquire more than half of their masses in 5 Myr. The material ejection that occurred in impacts of bodies with the planets, which was not taken into account in the model, may enlarge the accumulation time for the planets. A relatively rapid growth of the bulk of the Martian mass can be explained by the formation of Mars' embryo (the mass of which is several times less than that of Mars) due to contraction of a rarefied condensation. For the mass ratio of the Earth's and lunar embryos equal to 81 (the same as that for the masses of the Earth and the Moon), the ratio of the probabilities for infalls of planetesimals onto the Earth's and lunar embryos did not exceed 54 for the considered variants of calculations; and it was highest for the embryos' masses approximately three times less than the present masses of these celestial bodies. Special features in the formation of the terrestrial planets can be explained even under a relatively gentle decrease of the semi-major axis of Jupiter's orbit due to ejection of planetesimals by Jupiter into hyperbolic orbits. In this modeling, it is not necessary to consider the migration of Jupiter to the orbit of Mars and back, as in the Grand Tack model, and sharp changes in the orbits of the giant planets falling into a resonance, as in the Nice model.

## INTRODUCTION

According to analytical estimates (Safronov, 1972; Vityazev et al., 1990; Wetherill, 1980), the formation of the Earth took approximately 100 Myr. The process of formation of the terrestrial planets were also analytically studied in many other papers (e.g., Shmidt, 1945; Gurevich and Lebedinskii, 1950; Safronov, 1954, 1958, 1960, 1975; Levin, 1964, 1978; Weiden-schilling, 1974; Ziglina and Safronov, 1976; Vityazev et al., 1978; Levin, 1978; Pechernikova and Vityazev, 1979, 1980; Safronov and Vityazev, 1985; Safronov, 1986; Safronov and Vitjazev, 1986; Lissauer, 1987, 1993; Vityazev, 1991; Ziglina, 1991, 1995; Greenberg et al., 1991). A large number of papers focused on numerical simulations of the formation of the terrestrial planets were published. The evolutionary model for disks of gravitating bodies which coagulate under collisions in the feeding zone of the terrestrial planets was considered by Cox (1978), Cox and Lewis (1980), Wetherill (1980, 1985, 1988a, 1998b), Ipatov (1981, 1982, 1987, 1992, 1993a, 2000), Lecar and Aarseth (1986), Wetherill and Stewart (1989), Beauge and Aarseth (1990), Chambers and Wetherill (1998), Chambers (2001, 2013), Raymond et al. (2004, 2006, 2009), O'Brien et al. (2006), Hansen (2009), Morishima et al. (2010), Morbidelli et al. (2012), Izidoro et al. (2014), Hoffmann et al. (2017), and Lykawka and Ito (2017). The characteristics and evolution of the terrestrial planets were analyzed by Marov (2017).

The first paper dealing with computer simulations of the accumulation of planets was published by Dole (1970), who postulated a concept of planetary embryos (in the algorithm, embryos are thrown into a disk and accumulate small objects, while the gravitational interactions are ignored). Kozlov and Eneev (1977) and Eneev and Kozlov (1979, 1981) studied the formation of protoplanets by coagulating the strongly rarefied gas−dust condensations that move along almost circular orbits. It was supposed in these papers that condensations had coagulated into giant rarefied protoplanets with masses equal to those of the present planets before they contracted to the density of solid bodies. The first studies on the evolution of rings of solid bodies, which simulated the gravitational inter-action between the bodies of the disk and considered two bodies to coagulate when the distance between their centers of masses is equal to the sum of their radii (rather than the radius of a large conventional sphere), were based on a two-dimensional model and appeared in 1978 (Ipatov, 1978; Cox, 1978). A three-dimensional model of evolution of these disks was considered for the first time by Wetherill (1980). Cox and Lewis (1980) and Wetherill (1980) assumed a number of initial bodies to be 100. In these papers, the gravitational influence of bodies was accounted for with the method of spheres (Cox et al., 1978).

Ipatov (1981, 1982, 1987, 1993a, 2000) considered the evolution of disks of gravitating bodies which coag-ulate under collisions in the feeding zone of the terrestrial planets. The mutual gravitational influence of bodies was taken into account with the method of action spheres, i.e., beyond the action spheres the bodies are moving about the Sun along the undisturbed Keplerian orbits, while the relative motion inside the action spheres is considered within the two-body problem. The distance of initial bodies to the Sun varied from 0.36 (or 0.4) to 1.2 AU. The total mass of the bodies was $1.87 m_E$, where $m_E$ is the Earth's mass. Simulations of the evolution of two-dimensional disks for the case of almost circular initial orbits (Ipatov, 1981, 1982) yielded a number of formed planets larger than four—the number of actually existing terrestrial planets—while the required number of planets was produced only under initial eccentricities of 0.35. In simulations of the evolution of three-dimensional disks, a real number of the planets may be obtained (Ipatov, 1982, 1987, 1993a, 2000). For example, four planets with masses larger than $0.046 m_E$ were formed in the evolutionary scenarios considered by Ipatov (2000). Ipatov (1982, 1987, 1993a, 2000) analyzed up to 1000 initial bodies in each of the disks. In calculations of the evolution of three-dimensional disks with the method of action spheres, the initial eccentricities $e_0$ were 0.02. It was shown that, for the disks under consideration, the values of eccentricities close to 0.02 are rather easily achieved, if the gravitational influence of bodies is taken into account at distances larger than the radii of action spheres. The bodies initially located at different distances from the Sun were distributed over semi-major axes and eccentricities at different stages of the disk evolution. In the course of the evolu-tion of three-dimensional disks, the averaged orbital eccentricity $e_{av}$ of the bodies exceeded 0.2, while it was larger than 0.4 in some scenarios at some times. For example, in the scenario presented in Fig. 6.1 of the paper by Ipatov (2000), 960 initial bodies and $e_0 = 0.02$ yielded $e_{av} = 0.09$, 0.20, and 0.35 for 500, 250, and 100 bodies in the disk, respectively. In this result, the bodies near the disk edges (the semi-major axes are $a < 0.4$ AU and $a > 1.2$ AU) exhibit the larger averaged orbital eccentricities. In the considered cases of calculations, the evolution of orbits for some planets with masses of an order of the Mercurian or Martian masses yielded eccentricities close to those for orbits of these planets and the bodies with larger eccentricities were mainly formed at the annulus periphery.

From the calculation results for the evolution of disks of bodies, Ipatov (1982, 1987, 1993a, 2000) drew the following conclusions on the accumulation of the terrestrial planets. In a real scenario containing a very large number of initial bodies, the hypothesis that Mercury and Mars acquired high eccentricities under the gravitational interaction between the bodies only from the feeding zone of the terrestrial planets means that the other bodies with masses close to that of Mercury and Mars also had high eccentricities in the course of evolution. Since the probability of ejecting bodies from the feeding zone of the terrestrial planets into hyperbolic orbits is low (10%), these bodies would most likely have collided with the embryos of the Earth and Venus. The increase in the eccentricities of Mercury and Mars (and the inclination of Mercury's orbit) could be partially caused by the gravitational influence of the bodies that f lew into the feeding zone of the terrestrial planets from the feeding zones of the giant planets. In this process, these bodies could not collide with bodies from the feeding zone of the terrestrial planets but only gravitationally disturb their orbits.

According to the estimates by Ipatov (2000, Chap. 6), the formation time for 80% of the mass of the largest planet (the Earth's analog) did not exceed 10 Myr, while the total time for the evolution of disks was approximately 100 Myr. Ipatov (1992) noted that the formation time for the most of the mass of planets, about 1–10 Myr, was obtained in consideration of the deterministic choice of pairs of colliding bodies in the disk (to model the encounter, a pair of bodies with the shortest time before the encounter was chosen). When the probabilistic method is used to choose the pairs of encountering bodies (a pair of bodies is chosen in proportion to the probability of their encounter) and the evolution of a disk of gravitating bodies is modeled with the method of action spheres, the formation time for the main mass of the planets turns out to be almost an order of magnitude larger than that obtained with the deterministic method. The deterministic method (Ipatov, 1993b, 2000) better reproduces the real evolution of disks of bodies than the probabilistic method used by Ipatov (1982, 1987) and yields estimates of the evolution time closer to those obtained later with numerical integration of the equations of motion (these papers are reviewed below).

The time during which the number of bodies in the disk decreases from $N_0$ to $N$, usually was approximately two times less than that taken by the decrease in the number of bodies in the disk from $N_0$ to $N/2$. It is the later stages of accumulation of planets that took most of the time of evolution of the considered disks. Because of this, Ipatov (1988, 2000) concluded that the entire duration of the disk evolution for $N_0 = 10^{12}$ is almost the same as that for $N_0 = 10^3$; however, accounting for the fragmentation of bodies under collisions may enlarge the time required to form the main mass of the planets.

The formation and migration of the giant planets are closely connected with the accumulation of the terrestrial planets. Planetesimals from the feeding zone of the giant planets, which acquired orbits with small perihelion distances in the course of evolution, disturbed the orbits of planetesimals and planetary embryos in the feeding zone of the terrestrial planets and bodies from the asteroid belt and collided with them. Changes in the orbits of Jupiter and Saturn induced changes in the resonance positions and contributed to sweeping the asteroid belt zone, some bodies from which could penetrate into the feeding zone of the terrestrial planets. Consequently, when studying the accumulation of the terrestrial planets, one should take into account the influence of the forming giant planets and the bodies from their feeding zone.

Ipatov (1993a, 2000) analyzed the evolution of the disk that initially contains the terrestrial planets, Jupiter, Saturn, 750 identical bodies of a total mass of 150$m_E$ at a distance $R$ ranging from 8 to 32 AU from the Sun, and 150 small bodies at $R$ between 2 and 4 AU. During the evolution of this disk, small bodies were swept away from the asteroid belt, while some massive bodies had highly eccentric orbits with semi-major axes smaller than 2 AU. These bodies completely penetrated into the feeding zone of the terrestrial planets. Simulations of the evolution of disks containing the same initial bodies, but also Jupiter and Saturn with the present masses and embryos of Uranus and Neptune with masses of 10$m_E$ on almost circular initial orbits yielded analogous results. The initial values of semi-major axes of orbits of these giant planets were 5.5, 6.5, 8, and 10 AU, respectively. In the course of this evolution, Uranus and Neptune acquired orbits close to their present orbits. For the first time, simulations of the migration of the giant planets of this kind had been presented by Ipatov (1991a, 1991b, 1993a) long before the studies on the Nice model appeared (Gomes et al., 2005; Morbidelli et al., 2005; Tsiganis et al., 2005 and more recent papers). Later, analogous calculations were carried out by Thommes et al. (1999), but they used the symplectic method for inte-gration. Zharkov and Kozenko (1990) and Zharkov (1993) were the first who suggested that the embryos of Uranus and Neptune were formed near the orbit of Saturn; this idea was based on the analysis of the com-position of these giant planets. Zharkov and Kozenko came to conclusion that the embryos of Uranus and Neptune had acquired hydrogen envelopes with masses of approximately (1–1.5)$m_E$ in the growth zone of Jupiter and Saturn before gas dissipated from the protoplanetary disk. The increase in the semi-major axes of the orbits of Saturn, Uranus, and Neptune and the decrease in the semi-major axis of the orbit of Jupiter were obtained by Fernandez and Ip (1984). However, due to limitations in the algorithm used in that paper, the ejection of planetesimals into hyperbolic orbits and the changes in the semi-major axes of orbits of the giant planets were relatively small. Specifically, when accounting for gravitational interactions of bodies with planets, Fernandez and Ip (1984) used the spheres which were substantially smaller than the action spheres and modeled the infall of bodies onto the planet for the case when the distance between them amounts to several (up to 8) radii of the planet. As distinct from Ipatov (1991a, 1993a), Fernandez and Ip (1984) ignored the mutual gravitational influence of planetesimals. In the papers by Ipatov (1991a, 1993a, 2000), the mass of planetesimals ejected from the feeding zone of the giant planets into hyperbolic orbits was an order of magnitude higher than that of planetesimals incorporated into planets.

In contrast to the Nice model, Ipatov (1991a, 1991b, 1993a, 2000) considered the migration of embryos of Uranus and Neptune for the case when the giant planets did not enter into resonances and the total mass of planetesimals in the feeding zones of Uranus and Neptune was larger. In the scenarios, the latter was varied from $135m_E$ to $180m_E$. More than 80% of planetesimals were ejected into hyperbolic orbits. Ipatov (2000) concluded that the disk of planetesimals with a mass of $100m_E$ is enough for the embryos of Uranus and Neptune to migrate to the present orbits. This mass is smaller if the larger (than those in the calculations) semi-major axes of the initial orbits of embryos of Uranus and Neptune are considered (in the calculations, they were 8 and 10 AU, respectively). In simulations by Ipatov (1993a, 2000), the main changes in the orbital elements of embryos of the giants took place for a time not exceeding 10 Myr, though some bodies could fall onto these embryos in approximately billions of years. If the main part of the disk mass was in small bodies, the migration time for planetary embryos could be larger than that in the calculations (the masses of initial bodies was assumed at $0.2m_E$). In addition to simulating the migration of the giant planets which initially were on circular or almost circular orbits, Ipatov (1991a, 1993a, 2000) considered the scenario proposed in 1990 by Zharkov for large (0.75–0.82) initial eccentricities of massive ($10m_E$) embryos of Uranus and Neptune. In this case the orbital eccentricities of these embryos decreased under interaction of the embryos with planetesimals, and the embryos could also get to the present orbits of Uranus and Neptune, if the initial perihelia of their orbits were beyond the orbit of Saturn (if the perihelion distances were smaller, these embryos were most often ejected into hyperbolic orbits). However, it appears unlikely that the embryos of Uranus and Neptune acquired these eccentric orbits with perihelia beyond Saturn's orbit.

It is believed (Cameron and Pine, 1973; Torbett and Smoluchowski, 1980; Safronov and Ziglina, 1991; Safronov, 1991) that the resonance scanning connected with a change in the semi-major axis of Jupiter's orbit could be, in addition to the inf luence of bodies from the feeding zones of the giant planets, one of the causes of sweeping bodies from the asteroid belt. In the simulation scenarios considered by Ipatov (1993a; 2000), the semi-major axis of Jupiter's orbit decreased approximately by $0.005m_{un}^o/m_E$ (expressed in astronomical units), while the semi-major axis of Saturn's orbit increased by $(0.01–0.03)m_{un}^o/m_E$, where $m_{un}^o$ is the total initial mass of bodies in the feeding zone of Uranus and Neptune. Thus, if $m_{un}^o/m_E{\geq}100$, the shifting resonances covered a substantial part of the asteroid belt. To transport the embryos of Uranus and Neptune from the vicinity of Saturn's orbit to the present orbits, the ratio $m_{un}^o/m_E{\approx}100$ was required (Ipatov, 1991a, 1991b, 1993a, 2000). In the course of evolution of the disks considered in the cited papers, about 1% of orbits of the bodies initially located in the feeding zones of Uranus and Neptune crossed the Earth's orbit at some evolution stages. During the evolution of these disks, the mass of bodies ejected into hyperbolic orbits was an order of magnitude larger than the mass of bodies incorporated within the planet.

Ipatov (1993a, 2000) also studied the evolution of disks initially composed of the terrestrial planets, Jupiter, Saturn, 250 planetesimals with a total mass $m_{js}^o = 10m_E$ and semi-major axes of initial orbits ranging from 5 to 10 AU, and 250 asteroid-like bodies with semi-major axes of initial orbits between 2 and 5 AU. In this simulation scenario, the evolution resulted in decreasing the semi-major axes of Jupiter and Saturn by $0.005m_{js}^o/m_E$ and $0.01m_{js}^o/m_E$ AU, respectively. In other words, the dependences of the change in the semi-major axis of Jupiter on $m_{un}^o$ and $m_{js}^o$ were almost the same. Consequently, the changes in the semi-major axis of Jupiter were mainly dependent on the total mass of planetesimals in the feeding zone of the giant planets rather than its distribution over distances in this zone. In the calculations, to consider the influence of smaller planetesimals on asteroids, we assumed the $m_{js}^o$ values to be substantially smaller than the actual total mass of planetesimals in the zone of Jupiter and Saturn. The evolution of similar disks composed of asteroids and massive bodies in the zone of Jupiter and Saturn was also analyzed by Wetherill (1989). Ipatov (1993a, 2000) and Wetherill (1989) obtained the growth of the mean orbital eccentricities of asteroids to the values not smaller than those in the present asteroid belt. Most asteroids were ejected into hyperbolic orbits. In the simulations by Ipatov (1993a, 2000), 5 and 2.5% of asteroids fell onto Venus and the Earth, respectively. In these calculations, Mercury and Mars even left the Solar System, which was apparently caused by that the assumed masses of the bodies ($0.04m_E$) substantially exceeded the mean masses of real planetesimals in the feeding zones of Jupiter and Saturn.

At the beginning, the semi-major axis of Jupiter's orbit decreased by $0.005m_{js}^o/m_E$ AU for several million years, since Jupiter ejected the bodies from the feeding zone of Jupiter and Saturn; later, this process was slower, by $0.005m_{un}^o/m_E$, due to ejection of the bodies initially located beyond Saturn's orbit. During this, the positions of resonances were changing and some bodies penetrated the asteroid belt zone and the feeding zone of the terrestrial planets.

The portion of planetesimals ejected from the feeding zones of the giant planets was also estimated by Ipatov (2019) in the study of migration of planetesimals initially located at different distances from the Sun. In these calculations, the evolution of orbits of planetesimals under the influence of planets was mod-eled by numerical integration of the equations of motion. The probability of a collision with Uranus or Neptune for a planetesimal that initially was beyond Jupiter's orbit did not exceed 0.015 and were not more than a few thousandths in most simulations.

Consequently, if the total mass of planetesimals beyond Saturn's orbit was less than $200m_E$, massive embryos of Uranus and Neptune which had migrated from the vicinity of Saturn's orbit to the present orbits built up their masses by less than $2m_E$. The collision probability of a planetesimal with Jupiter did not exceed 0.05 in most simulations, and that with Saturn was several times smaller. According to Gudkova and Zharkov (1999) and Zharkov (2003), the mass of a silicate component in Jupiter is $(15–20)m_E$. Planetesimals in the feeding zones of Uranus and Neptune also contained ices along with silicates. Because of this, if the total mass of planetesimals beyond Saturn's orbit was smaller than $200m_E$, the increase in the mass of Jupiter's silicate component due to these planetesimals did not exceed several masses of the Earth. In the composition of Saturn, the mass of a solid component is larger than that for Jupiter (Zharkov, 1991, 2013; Zharkov and Gudkova, 2019). Consequently, the total mass smaller than $200m_E$ for planetesimals beyond Saturn's orbit does not contradict the composition of the giant planets. If the ratio of the masses of dust (rocks and ices) to gas is 0.015 (Lodders, 2003; Ziglina and Makalkin, 2016), then the total mass of planetesimals equal to $200m_E$ corresponds to the mass of a proto-planetary disk beyond Saturn's orbit of $0.04M_S$, where $M_S$ is the solar mass (in this case, the total mass of the disk is $0.06M_S$). Approximately the same values of the mass of a protoplanetary disk, $(0.04–0.1)M_S$, were considered in many papers on cosmogony (Safronov, 1972; Safronov and Vityazev, 1985; Ruzmaikina and Maeva, 1986; Makalkin and Artyushkova, 2017).

With the emergence of more powerful computer processors, for computer simulations of disks of bodies coagulating at collisions, corresponding to the feeding zone of the terrestrial planets, the mutual gravitational inf luence of bodies began to be taken into account with the use of numerical integration of the equations of motion (Chambers and Wetherill, 1998; Chambers, 2001, 2013; Raymond et al., 2004, 2006, 2009; Han-sen, 2009; Morishima et al., 2010; Izidoro et al., 2014; Hoffmann et al., 2017; Lykawka and Ito, 2017). In calculations by Chambers and Wetherill (1998), the highest number of planetary embryos was 56 for each of the cases; all their calculations took almost three years of processing time. Recently, the number of bodies in calculations has reached several thousands. For example, Lykawka and Ito (2017) considered 6000 planetesimals.

Raymond et al. (2004) considered different orbits and masses for Jupiter. In their simulations, the terres-trial planets approached half of their final masses for the first 10–20 Myr, though some bodies fell onto them for over 100 Myr. Raymond et al. (2006, 2009) considered 1000–2000 planetesimals in the primordial disk, i.e., 5–10 times larger than those in the previous papers, where the mutual gravitational influence was taken into account by numerical integration of the equations of motion. The primordial disk $9.9m_E$ in mass extended to 5 AU. Over a billion years, more than 99% of the asteroid belt was swept away, since planetesimals fell into resonances with Jupiter due to their mutual gravitational influence and the inf luence of embryos.

Morishima et al. (2010) and Hoffmann et al. (2017) took into account the influence of gas. Morishima et al. (2010) noted that, for the present value of the orbital eccentricity of Jupiter, most bodies of the aster-oid belt are swept out from it due to the secular reso-nance motion. Hoffmann et al. (2017) noted that, while gas was dissipating, secular resonances ($\upsilon_5$, $\upsilon_6$, $\upsilon_{15}$, and $\upsilon_{16}$) with Jupiter and Saturn were moving inward, pushing planetesimals ahead. In 3 Myr, the gaseous disk became smaller in mass by 20 times and did not influence the migration dynamically. Ohtsuki et al. (1988) took into account the drag effect of gas and obtained the growth of the Earth in 10 Myr.

Kokubo and Ida (2000) considered the evolution of narrow annuluses of planetesimals 0.02 and 0.092 AU wide at a distance of 1 AU from the Sun. The gas drag effect on the motion of planetesimals was included in the simulation. At a distance of 1 AU from the Sun, the formation time turned out to be 0.5 Myr for protoplan-ets with masses of $\sim 10^{26}$ g. The masses of embryos did not exceed $0.16m_E$ ($m_E \approx 6 \times 10^{27}$ g is the Earth's mass). The evolution of analogous narrow annuluses, though with the gas effect ignored, was analyzed by Kokubo and Ida (1998). In that paper, the authors showed that the distance between the orbits of the form-ing embryos was $(5–10)r_H$, where $r_H$ is the Hill radius of embryos. The results by Weidenschiling et al. (1997) and Kokubo and Ida (2000) suggest that, in 1 Myr, the main portion of the annulus mass was in the bodies with masses larger than few $10^{26}$ g. In semianalytic models (Chambers, 2006), the formation of an embryo $0.1m_E$ in mass at a distance of 1 AU and an embryo $10m_E$ in mass at a distance of 5 AU took 0.1 and 1 Myr, respectively.

In the simulations by Lykawka and Ito (2017), most of the mass incorporated into analogs of Mercury came from a zone of 0.2 to 1.5 AU for 10 Myr, while the remainder came later from a zone stretching out to 3 AU (the disk with a mass of $7m_E$ at a distance of 0.2 to 3.8 AU from the Sun was considered). The mean mass of these analogs was of an order of $0.2m_E$, i.e., exceeded the mass of Mercury. The semi-major axes of orbits of these analogs were close to 0.27–0.34 AU, while the eccentricities and inclinations were small. Raymond and Izidoro (2017) believed that the pri-mordial asteroid belt could be empty. Hansen (2009) studied the evolution of a narrower (at a distance between 0.7 and 1.0 AU from the Sun) annulus than that considered by the other authors. In his calcula-tions, the analogs of the Earth and Mars accumulated most of their mass in 10 Myr. According to Kokubo and Genda (2010), only half of collisions of planetesi-mals and embryos resulted in accretion.

Morbidelli et al. (2010) considered the evolution of orbits of asteroids during a sharp change in Jupiter's orbit that led to a sharp change in the positions of resonances. The probability of collisions of asteroids with the Moon was found to be $4 \times 10^{-5}$, while this quantity for the Earth was 20 times higher. Clement et al. (2018, 2019) analyzed the formation of the terrestrial planets during this instability of the orbits of the giant planets. They found that, if this instability (within the Nice model) occurred during a span of 1–10 Myr after the gaseous disk dissipation, then the model successfully explains the formation of Mars and the asteroid belt.

At the early stages of the Solar System evolution, gas played an important role. When studying the accu-mulation of bodies in a zone between 0.5 and 4 AU, Hoffmann et al. (2017) supposed that the projected density of gas exponentially decreased with time: $\Sigma_{gas}(r, t) = \Sigma_{gas,0}(r/1 \; AU)^{-1}\exp(-t/\tau)$. In these calculations, it was assumed that $\tau = 1$ Myr, while $\Sigma_{gas,0} = 2000$ g/cm$^2$. Only 1% of gas remained in 4.6 Myr. The accumulation of bodies in the zone of the terrestrial planets was also influenced by the gravity of Jupiter and other bodies from its feeding zone. The papers on the formation of Jupiter were reviewed by D'Angelo and Lissauer (2018). When planetesimals were accu-mulated, Jupiter's embryo at a distance of 5.2 AU could reach a mass of $3m_E$ in 0.1 Myr; however, after that, not many planetesimals remained in the vicinity of its orbit. If Jupiter's embryo was growing in the gaseous medium by accumulation of solid objects 1 cm to 1 m in size, its mass could increase to $10m_E$ in a time less than several tens of thousands of years. When the mass of a solid component of Jupiter's embryo and the mass of gas from its closest vicinity reached some critical values, the stage of rapid gas accretion began, during which Jupiter's embryo built up its mass several fold for ~0.1 Myr. In some models, the total time for growing the mass of Jupiter from zero to approximately the present value is about 2 Myr (D'Angelo and Lissauer, 2018). However, having already reached a mass of approximately $10m_E$ in ~0.1 Myr, Jupiter's embryo could increase the orbital eccentricities of bodies from its feeding zone in such a way that they could reach the feeding zone of the terrestrial planets at perihelion. As the masses of planetesimals grew and the gas density in the feeding zone of Jupiter decreased, the mutual gravitational influence of planetesimals enhanced the capacity of some planetesimals to start crossing Jupiter's orbit and, after that, achieve small perihelion distances.

In some papers dealing with the formation of the terrestrial planets (Walsh et al., 2011; Morbidelli et al., 2011; Jacobson and Morbidelli, 2014; O'Brien et al., 2014; Rubie et al., 2015), the Grand Tack model was considered. In this model, Jupiter was first moving, under the presence of gas, toward the Sun, to 1.5 AU; then, after formation of massive Saturn, it started to move together with Saturn away from the Sun, being in the 2:3 resonance with Saturn. In the course of this migration, Jupiter swept the asteroid belt and diminished the amount of material in the feeding zone of Mars. This migration also explains the delivery of water to the forming terrestrial planets. During the migration of Saturn from the Sun, many bodies, which were beyond 6 AU from the Sun, migrated towards the Sun. Rubie et al. (2015) supposed that gas was present for 0.6 Myr after Jupiter and Saturn had acquired rather large masses; moreover, for the first 0.1 Myr, Jupiter and Saturn migrated inward, from 3.5 and 4.4 AU to 1.5 and 2 AU, respectively. Once Saturn's mass had increased from $10m_E$ to its present value, Jupiter and Saturn were migrating from the Sun for 0.5 Myr to a distance of 5.25 and 7 AU, respectively. Water was delivered to the Earth mainly after the accretion of 60–80% of its final mass. Dorofeeva and Makalkin (2004) noted that volatiles in the Earth's zone could be accumulated by parent bodies in only 1 Myr of the evolution of the preplanetary circumsolar disk, when its temperature fell below 700 K. The water-containing bodies were formed at a distance from the Sun larger than 6 AU. In different scenarios the time required for the Earth's analog to grow to $0.5m_E$ took the values from several to 20 Myr, while its collision with a body (0.03–0.16)$m_E$ in mass occurred 20–150 Myr after the disk evolution had started (O'Brien et al., 2014).

Drolshagena et al. (2017) noted that the amount of the material falling onto the Earth's atmosphere every day is approximately 30–180 t. For the objects smaller than 0.5 m, this estimate is 32 t per day; moreover, among objects smaller than few centimeters, particles with sizes from 10 μm to a millimeter are of prime importance in accumulation. Due to constant mutual collisions of planetesimals, the bulk mass of small objects and dust was not small during the accumulation of planets. The collision probability of a particle approximately 100 μm across with a planet could exceed by orders of magnitude the collision probability of a planetesimal with a planet for the same initial orbits of planetesimals and dust particles (Ipatov and Mather, 2006; Ipatov, 2010a). This is caused by smaller typical eccentricities (and, consequently, smaller relative velocities) of these particles (as compared to those of planetesimals), when they usually encountered a planet.

From the analysis of the ratio of lead isotopes, $^{207}Pb/^{206}Pb$, in zircon crystals contained in the material of the Martian meteorite NWA 7034, Bouvier et al. (2018) concluded that the core formation and the crystallization of a magma ocean on Mars was completed in less than 20 Myr after the formation of the Solar System. These results agree with those reported by Mezger et al. (2013) concerning the studies of the decay of a system of short-lived isotopes $^{182}Hf-^{182}W$, which point to an age not larger than 10 Myr after the formation of the Solar System. Thermal models presented by Elkins-Tanton (2008) suggest that the solid-ification history of Mars was completed during 10 Myr of accretion. Elkins-Tanton (2008) supposed that Mars apparently grew to approximately the present size in less than 5 Myr after the formation of cal-cium-aluminum inclusions (CAIs). Nimmo and Kleine (2007) came to conclusion that the ratio Hf/W for the Martian mantle is ~4 with an uncertainty of ~25%, which results in a range of 0 to 10 Myr for the formation time of the Martian core.

**Table 1.** Portions of the bodies which formed a planet and initially were in four zones (at 0.4–0.6, 0.6–0.8, 0.8–1.0, and 1.0–1.2 AU from the Sun)

| Primordial disk | 0.156 | 0.219 | 0.281 | 0.344 |
|---|---|---|---|---|
| Planets with a mass $m > 0.5m_E$ | 0.148–0.187 | 0.194–0.249 | 0.25–0.303 | 0.319–0.353 |
| Planets with a mass $m > 0.1m_E$ | 0.114–0.217 | 0.191–0.249 | 0.193–0.336 | 0.319–0.398 |

A limitation of less than 30 Myr for forming most of the mass of the Earth and the Moon was obtained by Kleine et al. (2002) and Yin et al. (2002) from the analysis of the ratios Hf/W. Williams and Sujoy (2019) analyzed the ratio $^{20}Ne/^{22}Ne$ and found that the presence of nebular neon requires the Earth's embryo reach a substantial mass in several millions of years in order to trap nebular gasses and dissolve them in a magma ocean.

As distinct from the above-cited papers, where relatively small periods are considered for forming the terrestrial planets, Galimov (2013) concluded from the analysis of the $^{182}Hf–^{184}W$ system that the formation of cores of the Earth and the Moon could not have started earlier than 50 Myr after the origin of the Solar System. From the analysis of the Rb–Sr system, Galimov reached the conclusion that, before the Moon was formed as a condensed body, it had to evolve in a medium with a higher ratio of Rb/Sr. Since the atomic weight of rubidium is large, it cannot escape from the lunar surface, but can escape only from the heated surface of small bodies and particles. Consequently, according to Galimov, for the first 50 Myr, the primordial lunar material was in a disperse phase, for example, in the form of a gas–dust condensation.

Different models for the formation of the Moon were discussed by Ipatov (2018). Below, the analysis is carried out within the multiimpact model, which considers the infall of a large number of bodies onto embryos of the Earth and the Moon and the growth of the lunar embryo mainly at the expense of the material ejected in collisions of planetesimals with the Earth's embryo.

This study is mainly focused on the mixing of planetesimal bodies in the feeding zone of the terrestrial planets, the estimates of a relative amount of planetesimals initially located at different distances from the Sun, and their infalls onto different forming terrestrial planets. While modeling the evolution of gravitating bodies that coagulate in collisions, Ipatov (1993a, 2000) divided the primordial disk, corresponding to the feeding zone of the terrestrial planets and containing the initially identical bodies, into four zones in dependence on the distance of the bodies from the Sun (0.4–0.6, 0.6–0.8, 0.8–1.0, and 1.0–1.2 AU). The initial number of bodies in the disk reached 1000. It was obtained that the bodies initially located at different distances from the Sun were incorporated into the Earth and Venus in similar proportions. Ipatov (2000) presents the following table, demonstrating the composition of formed planets for several variants of calculations (see Table 1). In Table 1, the portions of bodies that formed a planet and were initially located in the abovementioned zones are listed.

In the first line of Table 1, there are portions of the bodies which initially were in different zones (a sum of the portions is 1). Table 1 presents the data on all formed planets without any dependence on their distances to the Sun. The formed planet with a mass of Mercury (there are no planets of this mass in Table 1) was composed of the bodies coming from different zones, but a fraction of the bodies incorporated from a certain zone differed mainly twofold from the fraction of bodies in this zone in the total composition of the primordial disk.

Chambers (2013) considered the initial disk containing 14 embryos with masses of $0.093m_E$ and 140 smaller bodies with masses of $0.0093m_E$. In the cited paper, analogously to the paper by Chambers (2001), the author analyzed the composition of the formed planets which initially were at 0.4–0.7, 0.7–1.1, 1.1–1.5, and 1.5–2.0 AU from the Sun. The composition of large planets formed at a distance of approximately 1 AU from the Sun could differ more than twofold in different simulation scenarios, even for their main component that came from the region at a distance of 0.7–1.1 AU from the Sun. Probably, this was connected with the small number of initial objects considered. The planet that is similar to the Earth reached half of its final mass in approximately 20 Myr. O'Brien (2006) considered the composition of formed planets from both the same zones, as those in the papers by Chambers (2001, 2013), and the wider zones with distances from the Sun in ranges of 0.3–2, 2–2.5, 2.5–3, and 3–4 AU. The initial number of bodies was roughly 1000. The planets with semi-major axes smaller than 2 AU were mainly formed from the material of the zone at a distance of 0.3 to 2.0 AU from the Sun. For the formed terrestrial planets, the average fraction of the material from a zone beyond 2.5 AU from the Sun turned out to be 15 and 0.3% for circular and eccentric orbits of Jupiter and Saturn, respectively. Circular orbits of Jupiter and Saturn were considered in the Nice model.

In this paper, we first discuss the initial data and simulation algorithms for the model of the migration of planetesimals-bodies and the algorithms to calculate the collision probability of the bodies with the growing terrestrial planets. In the next sections, the results of calculations are reported; and the probabilities of infalls of planetesimals, which initially were in different regions of the feeding zone of the terrestrial planets, onto the planets and their embryos are discussed. Further, we consider the probabilities of infalls of planetesimals onto the Sun and the giant planets, the probabilities of ejection of planetesimals into hyperbolic orbits, and the probabilities of their collisions with the lunar embryo. Finally, the formation of the terrestrial planets is discussed.

## INITIAL DATA AND ALGORITHMS TO SIMULATE THE MIGRATION OF PLANETESIMALS AND THE PROBABILITIES OF THEIR COLLISIONS WITH PLANETS

In this paper, we consider the planetesimals which initially were in a relatively narrow annulus and report the results of calculations of their migration under the gravitational influence of the planets or their embryos. In each of the simulation scenarios, the number of initial planetesimals was $N_0$=250. The initial values $a_0$ of semi-major axes $a$ of orbits of planetesimals were changed from $a_{0min}$ to $(a_{0min}+d_a)$ AU, and the number of planetesimals with $a_0$ was proportional to $a_0^{1,2}$. For the $(i+1)$th planetesimal, the value of $a_0$ was calculated with the formula $a_{0(i+1)}=(a_{0i}^2+[(a_{0min}+d_a)^2-a_{0min}^2]/N_0)^{1/2}$, where $a_{0i}$ is the value of $a_0$ for the $i$th planetesimal. The values of $a_{0min}$ were varied from 0.3 to 1.5 AU. For $a_{0min}$=1.5 AU, it was assumed that $d_a$=0.5 AU. In the other cases of calculations, $d_a$=0.2 AU. All of the distances below are expressed in astronomical units. As has been mentioned in the Introduction, Chambers (2001, 2013) and O'Brien (2006) considered the composition of the planets formed from the bodies which initially were at 0.4–0.7, 0.7–1.1, 1.1– 1.5, and 1.5–2.0 AU from the Sun. In the interval within 1.5 AU from the Sun, these authors considered three zones instead of six zones considered here.

The initial eccentricities $e_0$ of orbits of planetesimals were assumed at 0.05 and 0.3 in different simulation scenarios. The initial inclinations of planetesimals were $i_0 = e_0/2$ rad. It was found by Ipatov (1982, 1987, 1993a, 2000) that, due to mutual gravitational influence of planetesimals, the mean orbital eccentricity of planetesimals in the feeding zone of the terrestrial planets could exceed 0.2 in the course of evolution.

In a series of calculations called MeN, we considered the planetesimals which initially were located in a relatively narrow annulus and modeled their migration under the gravitational influence of all planets (from Mercury to Neptune). The masses and orbital elements of the planets were equal to the present values. In the MeN$_{03}$ calculation series, in contrast to the MeN series, the masses of embryos of the terrestrial planets were assumed to be $m_{rel} = 0.3$ of the present masses of these planets, while the masses and orbital elements of the giant planets were equal to their present values. In the MeN$_{01}$ calculation series, we considered the embryos of the terrestrial planets with masses $m_{rel} = 0.1$ of the present masses of these planets, which were moving along their current orbits, and Jupiter and Saturn with their current masses and orbits. In the MeS$_{01}$ series, Uranus and Neptune were ignored, since these planets were unlikely to have reached their present masses and orbits at the time when the masses of embryos of the terrestrial planets were small. The gravitational influence of Uranus and Neptune on the migration of planetesimals was weak for planetesimals in the zone of the terrestrial planets.

To model the migration of planetesimals, a symplectic integrator of the Swift integration package (Levison and Duncan, 1994) was used. In this integration procedure, the collisions of planetesimals with planets were not modeled (i.e., planetesimals and planets were considered as mass points), but planetes-imals were removed from the integration when they collided with the Sun or moved away from the Sun to a distance greater than 2000 AU.

The orbital elements of migrated planetesimals were saved in the computer memory with a step of 500 yr. Based on the arrays of the orbital elements, similar to simulations by Ipatov and Mather (2003, 2004a, 2004b) and Marov and Ipatov (2018), the probabilities of collisions of planetesimals with the planets, the Moon, and their embryos were calculated for time interval $T$. In this procedure, based on these arrays of the orbital elements of migrated planetesimals, the probabilities were calculated not only for collisions between the planetesimals and the planetary embryos and the Moon, which were considered in numerical integration of the equations of motion in the analysis of planetesimals' migration, but also for collisions between the planetesimals and the embryos of the other masses (though the embryos of the other masses were ignored in the integration). This kind of approach to the analysis of the growth of planetary embryos at the expense of planetesimals which initially were at different distances from the Sun has never been used before. As distinct from the earlier simulations of the evolution of disks of bodies coagulating under collisions, this approach enables us to calculate more accurately the probabilities of collisions of planetesimals with planetary embryos for some evolutionary stages.

In calculations of the probability $p_{dts}$ of the encounters of a planetesimal with a planet to the radius $r_s$ of the sphere under consideration (usually, this is an action sphere of a planet with a mass $m_{pl}$ and a radius $r_s \approx R(m_{pl}/M_S)^{2/5}$, where $M_S$ is the solar mass) for the time $d_t$ in the three-dimensional model, the following formulas were used (Ipatov, 1988, 2000, Chap. 4, §2): $p_{dts} = d_t/T_3$, where $T_3 = 2\pi^2 k_p T_s k_v$, $iR^2/(r_\Sigma^2 k_{fi})$ is the characteristic time of an encounter, $i$ is the angle between orbital planes of the encountering celestial objects (expressed in radians), $R$ is the distance of an encounter point of the objects to the Sun, $k_{fi}$ is the sum of angles (expressed in radians) with a vertex on the Sun, within which the distance between the projections of orbits (along a ray radiating from the Sun) is smaller than $r_s$ (this sum is different for different orbits (Ipatov, 2000, Fig. 4.1), $T_s$ is the synodic rotation period, $k_p = P_2/P_1$, $P_2 > P_1$, where $P_i$ is the rotation period of the $i$th object (a planetesimal or a planet) about the Sun, and $k_v = (2a/R - 1)^{1/2}$, where $a$ the semi-major axis of the planetesimal's orbit (the coefficient $k_v$ was introduced by Ipatov and Mather (2004a) to account for the dependence of the encounter velocity on the position of a planetesimal on the eccentric orbit). The collision probability for the objects that entered the action sphere was assumed to be $p_{dtc} = (r_\Sigma/r_s)^2(1 + (v_{par}/v_{rel})^2)$, where $v_{par} = (2Gm_\Sigma/r_\Sigma)^{1/2}$ is the parabolic velocity, $v_{rel}$ is the relative velocity of the objects separated by the distance $r_s$, $r_\Sigma$ is a sum of the radii of colliding objects with a total mass $m_\Sigma$, and $G$ is the gravitational constant. If $i$ is small, the other formulas were used in the algorithm. The algorithms (and their basis) to calculate $k_{fi}$ and the characteristic time between the collisions of objects are presented by Ipatov in Appendix 3 of the O-1211 report of the Keldysh Institute of Applied Mathemat-ics of the Russian Academy of Sciences for 1985 (pp. 86–130). The collision probability $p_{dt}$ for a planetesimal and a planet for the time $d_t$ is $p_{dts} p_{dtc}$. The values of $p_{dt}$ were summed for the dynamical lifetime of a planetesimal.

Our earlier studies (e.g., Ipatov and Mather, 2003, 2004a, 2004b, 2006, 2007; Ipatov, 2010a) of the migration of bodies in the Solar System and the delivery of water and volatiles to the terrestrial planets were first based on our numerical simulation results for the migration of tens of thousands of small bodies and dust particles, which started from these bodies, under the gravitational inf luence of all planets for the case when the initial orbits of bodies were close to the orbits of known comets, while the masses and orbital elements of the planets were equal to their present values. Marov and Ipatov (2018) simulated the migration of planetesimals to the terrestrial planet zone from the feeding zone of Jupiter and Saturn and considered the delivery of water and volatiles to the terrestrial planets and the Moon. Below, we consider the migration of planetesimals to the same planets and the Moon from the feeding zone of the terrestrial planets. In our previous calculations, we used the Bulirsh–Stoer method (BULSTO; Bulirsh and Stoer, 1966) and the symplectic integration method, which yielded almost the same results (Ipatov and Mather, 2004a, 2004b). Because of this, new calculations were carried out only with a faster symplectic method; and their results are presented below. For some series of simulations, where the initial orbits of bodies were close to the orbit of some comet, the values of the probability of a collision $p_E$ of the body with the Earth could differ almost by 100 times for different comets. Among almost 30000 objects, whose initial orbits crossed Jupiter's orbit (Jupiter-crossing objects, JCOs), several objects in the course of evolution acquired orbits lying entirely within Jupiter's orbit; and they were moving along these orbits for millions or even hundreds of millions of years. The collision probability of such an object with the terrestrial planet could be higher than the total probability for thousands of other objects, the initial orbits of which were almost the same, but they did not cross the Earth's orbit for a long time. To understand how the probabilities of collisions of plan-etesimals with the terrestrial planets and the Moon may depend on the initial orientation of the orbits of planetesimals from the feeding zone of the terrestrial planets (under the same orbital elements), we consider below the calculation scenarios differing only by the orientation of the initial orbits of planetesimals and the integration step. As will be shown below, the values of $p_E$ for these planetesimals (in contrast to the earlier consideration of bodies which came to the Earth from beyond Jupiter's orbit) do not strongly depend on the initial orientations of orbits and the integration step.

Tables 2–4 present the probabilities of collisions of planetesimals with the planets, the Moon, and their embryos for several values of the time interval $T$ ranging from 0.5 to 50 Myr. The results for calculation series MeS$_{01}$, MeN$_{03}$, and MeN are shown in Tables 2, 3, and 4, respectively. The probabilities of a collision of a planetesimal with the Earth, Venus, Mars, Mercury, Jupiter, Saturn, the Moon, and the Sun for the time $T$ are designated as $p_E$, $p_V$, $p_{Ma}$, $p_{Me}$, $p_J$, $p_S$, $p_M$, and $p_{Sun}$, respectively. Based on the arrays of orbital elements of the migrated planetesimals, we also calculated the probabilities $p_{E01}$, $p_{V01}$, $p_{Ma01}$, $p_{Me01}$, and $p_{M01}$ of collisions of a planetesimal with embryos of the terrestrial planets and the Moon, the masses of which were 10 times less than the present masses of the planets and the Moon. Similarly, we calculated the probabilities $p_{E03}$, $p_{V03}$, $p_{Ma03}$, $p_{Me03}$, and $p_{M03}$ of collisions of a planetesimal with embryos of the terrestrial planets and the Moon, the masses of which were 0.3 of the present masses of these celestial bodies. The notations $p_{Ma0}$, $p_{Me0}$, $p_{Ma03-0}$, $p_{Me03-0}$, $p_{Ma01-0}$, and $p_{Me01-0}$ are used for the probabilities of collisions of a planetesimal with Mars and Mercury and their embryos in the case of zero eccentricities of their orbits. Tables 2–4 also contain the ratios of the probabilities of collisions of a planetesimal with the planets, the Moon, and their embryos to those for this planetesimal and the Earth or its embryo. There are also ratios of the probability of a collision of a planetesimal with the Earth or its embryo to that for this planetesimal and the Moon or its embryo. The ratios of the probabilities of a collision of a planetesimal with the Moon and its embryo ($p_M/p_{M01}$ or $p_M/p_{M03}$) and the ratio $p_E/p_{E01}$ were also considered under the condition that the arrays of orbital elements of planetesimals are the same for the cases of the Moon and its embryo (or the Earth and its embryo). The probability of ejection of a planetesimal into a hyperbolic orbit for the time $T$ is designated as $p_{ej}$. The analogous calculation scenarios for the other initial orientations of orbits and the other integration step are marked by an asterisk in Tables 2–4. These calculation scenarios show a possible range of values for collision probabilities with the same orbits and masses of embryos and the same semi-major axes, eccentricities, and inclinations of orbits of planetesimals.

Shown in boldface in Tables 2–4 are the values of probabilities larger than 0.1 and the ratios of the probability of a collision between a planetesimal and a planet (or its embryo) to that for a planetesimal and the Earth (or its embryo), if this probability ratios are in a range from 0.5 to 2 relative to the ratio of the present masses of a planet and the Earth. The probability ratios in this range indicate that a substantial portion of the material analogous to that contained in the Earth's embryo was incorporated into the material of the planet, while the probabilities larger than 0.1 correspond to the calculation scenarios that contribute much to the growth of the embryos' masses. In Table 4, several cells for $T = 5$ Myr are empty and marked by "Abs", because the arrays of orbital elements of migrated planetesimals had been already deleted by the time when it was decided to consider also the probabilities of collisions with Mars and Mercury for their orbits with a different eccentricity. In some cells of Tables 2 and 3, there are no ratios of the probabilities of collisions of a planetesimal with embryos of planets to that of its collision with the Earth's embryo, because the latter is zero. These cells are marked by "Inf". The probability values larger than 1 indicate that an overwhelming majority of planetesimals fell onto the embryo in a time smaller than $T$. These large probabilities can be used to compare the collision probabilities for planetesimals and embryos of different planets. If $T$ is large, the probability ratios for collisions of planetesimals with different embryos correspond, as a rule, to large mean eccentricities of the orbits of planetesimals, which were increasing with time at $e_0 = 0.05$. Due to the mutual gravitational influence of planetesimals and the influence of planetesimals which penetrated into the feeding zone of the terrestrial planets from beyond Jupiter's orbit, the mean orbital eccentricities of real planetesimals in the feeding zone of the terrestrial planets could exceed the mean orbital eccentricities in the model, which takes into account only the gravitational inf luence of planets. Specifically, Ipatov (1982, 1987, 1993a, 2000) notes that, due to the mutual gravitational influence of planetesimals, their mean eccentricity reached 0.2–0.3 in the course of evolution. Because of this, Table 4 also presents the calculation results for $e_0 = 0.3$ rather than only for $e_0 = 0.05$. The larger values of $e_0$ usually yield the lower (sometimes, several times lower) probabilities of collisions of planetesimals with planets.

## PROBABILITIES OF COLLISIONS OF MIGRATED PLANETESIMALS WITH EMBRYOS OF THE TERRESTRIAL PLANETS

Based on the probabilities of collisions between planetesimals and embryos of the terrestrial planets, we may estimate the growth of these planetary embryos and the probabilities of collisions of planetesimals formed at different distances from the Sun with planetary embryos. In this section, these probabilities are considered within the model used for calculations. In these calculations, we did not take into account the mutual gravitational influence of planetesimals, which enlarged the orbital eccentricities and mixing of planetesimals in the feeding zone of the terrestrial planets. Consequently, the results presented in Tables 2–4 correspond to the minimal estimates of mixing of planetesimals in the feeding zone of the terrestrial planets.

When analyzing the results of calculations, it is necessary to take into account the fact that the embryos of the terrestrial planets were actually growing with different rates and, at certain times, their number could be larger than four, while the calculated masses of planetary embryos differed from the present masses of the terrestrial planets by the same factor. However, when the embryos were 10 times smaller than the present terrestrial planets in mass, it was assumed in the calculation scenarios that the embryos mainly accumulated the material only from the vicinity of their orbits; therefore, when the accumulation of relatively small embryos (with masses of 10% of the present masses of planets) is considered, it is not necessary to know exactly the masses of the other planetary embryos, if they were also small.

The larger the masses of planetary embryos, the faster their growth, if the other conditions are the same (however, the almost-formed planet could have already scooped out nearly all of the material from the vicinity of its orbit, while many planetesimals could still remain in the vicinity of the orbit of a smaller embryo). If the total mass of planetesimals in the feeding zone of the embryo is an order of magnitude larger than the embryo's mass, the embryo with a mass of 0.1 of the planetary mass built up its mass twofold by

**Table 2.** Probabilities of collisions of a planetesimal with the embryos of planets, the lunar embryo, and the Sun for the time interval $T$ (Myr) for the disks composed of 250 primordial planetesimals with the semi-major axes of orbits ranging from $a_{0min}$ to ($a_{0min} + d_a$) AU, the eccentricities $e_0 = 0.05$, and the inclinations $i_0 = e_0/2$ rad. The value $d_a = 0.2$ AU is assumed in all variants of the MeS$_{01}$ calculation series except that with $a_{0min} = 1.5$ AU, where $d_a = 0.5$ AU is assumed. **The ratio of masses of embryos of the terrestrial planets to the present masses of the planets was $m_{rel} = 0.1$.** The asterisked cases in Tables 2–4 differ only by the orientation of initial orbits of planetesimals and the integration step (their orbital elements and masses are the same as those not marked by an asterisk)

| $a_{0min}$, AU | 0.3 | 0.3 | 0.3* | 0.3* | 0.5 | 0.5 | 0.7 | 0.7 | 0.7 |
|---|---|---|---|---|---|---|---|---|---|
| $T$ | 1 | 5 | 0.5 | 5 | 5 | 20 | 1 | 2 | 5 |
| $p_{E01}$ | 0 | 0 | 0 | 0 | 0.024 | **0.196** | $1.8 \times 10^{-4}$ | 0.0022 | 0.023 |
| $p_{V01}$ | 0 | 0 | 0 | 0 | **0.358** | **0.995** | **0.217** | **0.40** | **0.819** |
| $p_{Ma01-0}$ | 0 | 0 | 0 | 0 | 0 | 0 | 0 | 0 | 0 |
| $p_{Ma01}$ | 0 | 0 | 0 | 0 | 0 | 0 | 0 | 0 | 0 |
| $p_{Me01-0}$ | 0.0649 | 0.0663 | 0.039 | 0.0539 | 0 | 0 | 0 | 0 | 0 |
| $p_{Me01}$ | 0.0317 | 0.0318 | 0.024 | 0.0298 | 0 | 0 | 0 | 0 | 0 |
| $p_{Sun}$ | 0.012 | 0.036 | 0 | 0.024 | 0.004 | 0.016 | 0 | 0 | 0 |
| $p_{ej}$ | 0 | 0 | 0 | 0 | 0 | 0 | 0 | 0.004 | 0 |
| $p_{V01}/p_{E01}$ | Inf | Inf | Inf | Inf | 15.0 | 5.1 | 1200 | 182 | 35.8 |
| $p_{Ma01-0}/p_{E01}$ | Inf | Inf | Inf | Inf | 0 | 0 | 0 | 0 | 0 |
| $p_{Ma01}/p_{E01}$ | Inf | Inf | Inf | Inf | 0 | 0 | 0 | 0 | 0 |
| $p_{Me01-0}/p_{E01}$ | Inf | Inf | Inf | Inf | 0 | 0 | 0 | 0 | 0 |
| $p_{Me01}/p_{E01}$ | Inf | Inf | Inf | Inf | 0 | 0 | 0 | 0 | 0 |
| $p_{J}/p_{E01}$ | Inf | Inf | Inf | Inf | 0 | 0 | 0 | 0 | 0 |
| $p_{Sun}/p_{E01}$ | Inf | Inf | Inf | Inf | 0.17 | 0.082 | 0 | 1.8 | 0 |
| $p_{E01}/p_{M01}$ | Inf | Inf | Inf | Inf | 19.8 | 18.1 | 25.6 | 16.7 | 23.5 |
| $p_{M}/p_{M01}$ | Inf | Inf | Inf | Inf | 4.59 | 4.69 | 5.31 | 4.55 | 5.12 |
| $p_{E}/p_{E01}$ | Inf | Inf | Inf | Inf | 8.49 | 8.19 | 15.8 | 9.78 | 9.44 |

| $a_{0min}$, AU | 0.7 | 0.9 | 0.9 | 0.9 | 0.9 | 1.1 | 1.1 | 1.1 |
|---|---|---|---|---|---|---|---|---|
| $T$ | 20 | 1 | 2 | 5 | 20 | 5 | 20 | 50 |
| $p_{E01}$ | **0.31** | **0.20** | **0.518** | **1.03** | **3.27** | 0.0010 | 0.0208 | 0.024 |
| $p_{V01}$ | **2.13** | 0 | 0.0007 | 0.013 | **0.208** | 0 | 0 | 0 |
| $p_{Ma01-0}$ | 0.0011 | 0 | $3.3 \times 10^{-5}$ | $4.4 \times 10^{-5}$ | 0.0034 | 0 | 0 | 0 |
| $p_{Ma01}$ | 0.0002 | 0 | $1.2 \times 10^{-5}$ | $1.8 \times 10^{-5}$ | $7.2 \times 10^{-4}$ | 0 | 0 | 0 |
| $p_{Me01-0}$ | 0.0013 | 0 | 0 | 0 | 0 | 0 | 0 | 0 |
| $p_{Me01}$ | $2.9 \times 10^{-4}$ | 0 | 0 | 0 | 0 | 0 | 0 | 0 |
| $p_{Sun}$ | 0.004 | 0 | 0 | 0 | 0.016 | 0 | 0 | 0 |
| $p_{ej}$ | 0 | 0 | 0 | 0 | 0 | 0 | 0 | 0 |
| $p_{V01}/p_{E01}$ | 6.9 | 0 | 0.0013 | 0.0126 | 0.0635 | 0 | 0 | 0 |
| $p_{Ma01-0}/p_{E01}$ | 0.0036 | 0 | $6.3 \times 10^{-5}$ | $4 \times 10^{-5}$ | 0.001 | 0 | 0 | 0 |
| $p_{Ma01}/p_{E01}$ | $6 \times 10^{-4}$ | 0 | $2.4 \times 10^{-5}$ | $2 \times 10^{-5}$ | $2.2 \times 10^{-4}$ | 0 | 0 | 0 |
| $p_{Me01-0}/p_{E01}$ | $9 \times 10^{-4}$ | 0 | 0 | 0 | 0 | 0 | 0 | 0 |
| $p_{Me01}/p_{E01}$ | 0.0009 | 0 | 0 | 0 | 0 | 0 | 0 | 0 |
| $p_{J}/p_{E01}$ | 0 | 0 | 0 | 0 | 0 | 0 | 0 | 0 |
| $p_{Sun}/p_{E01}$ | 0.013 | 0 | 0 | 0.17 | 0.082 | 0 | 0 | 0 |
| $p_{E01}/p_{M01}$ | 23.2 | 24.0 | 24.3 | 24.4 | 23.5 | 17.3 | 17.1 | 22.6 |
| $p_{M}/p_{M01}$ | 5.16 | 5.06 | 5.1 | 5.08 | 4.98 | 4.9 | 4.8 | 4.8 |
| $p_{E}/p_{E01}$ | 9.44 | 15.9 | 9.78 | 9.62 | 9.60 | 6.3 | 5.8 | 5.6 |

**Table 2.** (Contd.)

| $a_{0min}$, AU | 1.3 | 1.3 | 1.3 | 1.3 | 1.5 | 1.5 | 1.5 | 1.5 |
|---|---|---|---|---|---|---|---|---|
| $T$ | 1 | 5 | 20 | 50 | 1 | 5 | 20 | 50 |
| $p_{E01}$ | 0 | 0 | 0 | $2.1 \times 10^{-5}$ | 0 | $6.4 \times 10^{-5}$ | 0.0028 | 0.011 |
| $p_{V01}$ | 0 | 0 | 0 | 0 | 0 | 0 | 0 | $1.7 \times 10^{-4}$ |
| $p_{Ma01-0}$ | 0.0044 | 0.011 | 0.035 | 0.061 | 0.0015 | 0.0075 | 0.032 | 0.069 |
| $p_{Ma01}$ | 0.0045 | 0.015 | 0.053 | 0.073 | 0.0005 | $6.5 \times 10^{-4}$ | 0.0052 | 0.0134 |
| $p_{Me01-0}$ | 0 | 0 | 0 | 0 | 0 | 0 | 0 | 0 |
| $p_{Me01}$ | 0 | 0 | 0 | 0 | 0 | 0 | 0 | 0 |
| $p_{Sun}$ | 0 | 0 | 0 | 0 | 0 | 0 | 0 | 0 |
| $p_{ej}$ | 0 | 0 | 0 | 0 | 0 | 0 | 0 | 0 |
| $p_{V01}/p_{E01}$ | Inf | Inf | Inf | 0 | Inf | 0 | 0 | 0.015 |
| $p_{Ma01-0}/p_{E01}$ | Inf | Inf | Inf | 2330 | Inf | 117 | 11.4 | 6.3 |
| $p_{Ma01}/p_{E01}$ | Inf | Inf | Inf | 3420 | Inf | 10 | 1.9 | 1.2 |
| $p_{Me01-0}/p_{E01}$ | Inf | Inf | Inf | 0 | Inf | 0 | 0 | 0 |
| $p_{Me01}/p_{E01}$ | Inf | Inf | Inf | 0 | Inf | 0 | 0 | 0 |
| $p_J/p_{E01}$ | Inf | Inf | Inf | 0 | Inf | 0 | 0 | 0 |
| $p_{Sun}/p_{E01}$ | Inf | Inf | Inf | 0 | Inf | 0 | 0.082 | 0 |
| $p_{E01}/p_{M01}$ | Inf | Inf | Inf | 13.6 | Inf | 19.6 | 14.6 | 13.9 |
| $p_M/p_{M01}$ | Inf | Inf | Inf | 4.8 | Inf | 5.3 | 4.8 | 4.7 |
| $p_E/p_{E01}$ | Inf | Inf | Inf | 4.9 | Inf | 5.5 | 5.9 | 6.1 |

Designations: $p_{E01}$, $p_{V01}$, $p_{Ma01}$, $p_{Me01}$, and $p_{M01}$ are the probabilities of collisions of a planetesimal with embryos of the terrestrial planets—Earth, Venus, Mars, and Mercury—and the Moon, the masses of which were 10 times less than the present masses of the planets and the Moon. $p_{Ma01-0}$ and $p_{Me01-0}$ are the probabilities of collisions of a planetesimal with the embryos of Mars and Mercury with such masses for the case when the orbital eccentricities for these embryo were zero. $p_E$ and $p_M$ are the probabilities of collisions of a planetesimal with the Earth and the Moon of the present masses, respectively. $p_{Sun}$ and $p_J$ are the probabilities of collisions of a planetesimal with the Sun and Jupiter, respectively. $p_{ej}$ is the probability of ejection of a planetesimal into a hyperbolic orbit. All probabilities were calculated for the time interval $T$. The probability ratios corresponding to division by zero are marked by "Inf".

accumulating 10% of these planetesimals and, therefore, increased the probability of its collision with planetesimals moving along similar orbits. The growth in eccentricities and inclinations of orbits of planetesimals diminished the collision probability of one planetesimal with the embryo but could increase the number of planetesimals that may encounter the planetary embryo. If the calculations show some probability of collisions of planetesimals with such the embryo with a mass of 0.1 of the planetary mass, this probability for the embryos of a larger mass will be not smaller in most cases. If we suppose that, in the feeding zone of the terrestrial planets, the number of planetesimals with semi-major axes $a$ was proportional to $a^{1/2}$ (i.e., the number of planetesimals per unit area was proportional to $a^{-1/2}$) and take into account that the integral of $a^{1/2}$ is proportional to $a^{3/2}$, we find that the ratio of the number of planetesimals in seven zones with the boundary semi-major axes 0.3, 0.5, 0.7, 0.9, 1.1, 1.3, 1.5, and 2 AU will be proportional to 0.28, 0.35, 0.40, 0.45, 0.49, 0.53, and 1.49, respectively. The sum of the first five values is 1.97, which is close to the ratio of the total mass of the terrestrial planets to the Earth's mass, while the sum of the two last values is 2.02. In reality, a portion of two inner zones in the total mass of the disk located at a distance of 0.3–1.3 AU from the Sun could be smaller than these values; and the total mass of primordial planetesimals in this disk could exceed $2m_E$, if the infalls of planetesimals onto the Sun and their ejections to beyond the Martian orbit are accounted for. In this process, a relatively small growth of the terrestrial planets in mass took place at the expense of planetesimals that initially were at distances exceeding 2 AU.

The results of the MeS$_{01}$ calculation series showed that, if the masses of embryos of the terrestrial planets were approximately 0.1 of the planetary masses, the embryo grew mainly at the expense of planetesimals from its vicinity, and the embryos of the Earth and Venus grew faster than those of Mercury and Mars. For the MeS$_{01}$ calculation series, planetesimals from each of the considered zones could mainly collide with only one embryo, and the probabilities of collisions of planetesimals with the other embryos were zero or much smaller than that for this embryo.

For the MeS$_{01}$ calculation series and planetesimals with semi-major axes of initial orbits in a range of 0.9 to 1.1 AU (the total mass of these planetesimals could amount to $\geq 0.5 m_E$), the collision probability for a planetesimal and the Earth's embryo with a mass of $0.1 m_E$ was $p_{E01} = 0.2$, 0.5, and 1 for $T = 1$, 2, and 5 Myr, respectively. For planetesimals with semi-major axes

**Table 3.** Probabilities of collisions of a planetesimal with the embryos of planets, the lunar embryo, and the Sun for the time interval $T$ (Myr) for the disks composed of 250 primordial planetesimals with the semi-major axes of orbits ranging from $a_{0min}$ to ($a_{0min} + d_a$) AU, the eccentricities $e_0 = 0.05$, and the inclinations $i_0 = e_0/2$ rad. The value $d_a = 0.2$ AU is assumed in all variants of the MeN$_{03}$ calculation series except that with $a_{0min} = 1.5$ AU, where $d_a = 0.5$ AU is assumed. **The ratio of masses of embryos of the terrestrial planets to the present masses of the planets was $m_{rel} = 0.3$**

| $a_{0min}$, AU | 0.3 | 0.3 | 0.3 | 0.3 | 0.3 | 0.5 | 0.5 | 0.5 | 0.5 |
|---|---|---|---|---|---|---|---|---|---|
| $T$ | 1 | 2 | 5 | 10 | 20 | 1 | 2 | 5 | 10 |
| $p_{E03}$ | 0 | 0 | 0 | 0 | 0 | 0.0085 | 0.021 | 0.062 | **0.11** |
| $p_{V03}$ | 0.0002 | 0.003 | 0.010 | 0.019 | 0.094 | **0.367** | **0.610** | **1.05** | **1.42** |
| $p_{Ma03\text{-}0}$ | 0 | 0 | 0 | 0 | 0 | 0 | $4.4 \times 10^{-5}$ | $3.3 \times 10^{-4}$ | 0.0095 |
| $p_{Ma03}$ | 0 | 0 | 0 | 0 | 0 | 0 | $1.4 \times 10^{-5}$ | $4.5 \times 10^{-5}$ | $1.9 \times 10^{-4}$ |
| $p_{Me03\text{-}0}$ | 0.034 | 0.038 | 0.042 | 0.043 | 0.045 | 0.014 | 0.022 | 0.038 | 0.046 |
| $p_{Me03}$ | 0.048 | 0.051 | 0.053 | 0.053 | 0.054 | 0.003 | 0.0048 | 0.0086 | 0.011 |
| $p_{Sun}$ | 0.008 | 0.02 | 0.04 | **0.116** | **0.208** | 0 | 0 | 0.008 | 0.02 |
| $p_{ej}$ | 0 | 0 | 0 | 0.004 | 0.008 | 0 | 0 | 0 | 0 |
| $p_{V03}/p_{E03}$ | Inf | Inf | Inf | Inf | Inf | 43.1 | 29.5 | 16.9 | 12.8 |
| $p_{Ma03\text{-}0}/p_{E03}$ | Inf | Inf | Inf | Inf | Inf | 0 | 0.0021 | 0.0053 | 0.0085 |
| $p_{Ma03}/p_{E03}$ | Inf | Inf | Inf | Inf | Inf | 0 | 0.0007 | 0.0007 | 0.0018 |
| $p_{Me03\text{-}0}/p_{E03}$ | Inf | Inf | Inf | Inf | Inf | 1.61 | 1.07 | 0.615 | 0.415 |
| $p_{Me03}/p_{E03}$ | Inf | Inf | Inf | Inf | Inf | 0.32 | 0.236 | 0.138 | 0.096 |
| $p_J/p_{E03}$ | Inf | Inf | Inf | Inf | Inf | 0 | 0 | 0 | 0 |
| $p_{Sun}/p_{E03}$ | Inf | Inf | Inf | Inf | Inf | 0 | 0 | 0.13 | 0.18 |
| $p_{E03}/p_{M03}$ | Inf | Inf | Inf | Inf | Inf | 39.5 | 33.4 | 30.4 | 28.2 |
| $p_M/p_{M03}$ | Inf | Inf | Inf | Inf | Inf | 2.51 | 2.47 | 2.40 | 2.35 |

| $a_{0min}$, AU | 0.5 | 0.7 | 0.7 | 0.7 | 0.7 | 0.7* | 0.7* | 0.7* |
|---|---|---|---|---|---|---|---|---|
| $T$ | 20 | 1 | 2 | 5 | 10 | 1 | 2 | 5 |
| $p_{E03}$ | **0.21** | 0.074 | **0.22** | **0.81** | **1.93** | 0.061 | **0.228** | **1.12** |
| $p_{V03}$ | **2.02** | **0.24** | **0.44** | **0.97** | **1.84** | **0.26** | **0.497** | **1.60** |
| $p_{Ma03\text{-}0}$ | 0.012 | 0 | $1.5 \times 10^{-4}$ | 0.002 | 0.0080 | 0.036 | $4.5 \times 10^{-4}$ | 0.0037 |
| $p_{Ma03}$ | $4.9 \times 10^{-4}$ | 0 | $3.8 \times 10^{-5}$ | $6 \times 10^{-4}$ | 0.0018 | 0.0031 | $6.6 \times 10^{-5}$ | $7.8 \times 10^{-4}$ |
| $p_{Me03\text{-}0}$ | 0.053 | 0 | 0 | 0 | 0 | 0 | 0 | 0.0056 |
| $p_{Me03}$ | 0.013 | 0 | 0 | 0 | 0 | 0 | 0 | 0.0009 |
| $p_{Sun}$ | 0.056 | 0 | 0 | 0 | 0 | 0 | 0 | 0.008 |
| $p_{ej}$ | 0.004 | 0 | 0 | 0 | 0 | 0 | 0 | 0 |
| $p_{V03}/p_{E03}$ | 9.63 | 3.26 | 1.99 | **1.20** | **0.96** | 4.21 | 2.18 | **1.43** |
| $p_{Ma03\text{-}0}/p_{E03}$ | 0.0102 | 0 | $6.9 \times 10^{-4}$ | 0.003 | 0.0042 | 0.0023 | 0.0020 | 0.0033 |
| $p_{Ma03}/p_{E03}$ | 0.0023 | 0 | $1.7 \times 10^{-4}$ | $8 \times 10^{-4}$ | 0.0010 | 0.0002 | $2.9 \times 10^{-4}$ | 0.0007 |
| $p_{Me03\text{-}0}/p_{E03}$ | 0.25 | 0 | 0 | 0 | 0 | 0 | 0 | 0.0050 |
| $p_{Me03}/p_{E03}$ | 0.06 | 0 | 0 | 0 | 0 | 0 | 0 | 0.0008 |
| $p_J/p_{E03}$ | 0.016 | 0 | 0 | 0 | 0 | 0 | 0 | 0 |
| $p_{Sun}/p_{E03}$ | 0.27 | 0 | 0 | 0 | 0 | 0 | 0 | 0.007 |
| $p_{E03}/p_{M03}$ | 53.6 | 52.3 | 47.6 | 44.0 | 43.4 | 50.6 | 53.7 | 44.6 |
| $p_M/p_{M03}$ | 2.3 | 2.3 | 2.52 | 2.3 | 2.54 | 2.35 | 2.53 | 2.50 |

**Table 3.** (Contd.)

| $a_{0min}$, AU | 0.9 | 0.9 | 0.9 | 0.9 | 0.9 | 0.9 | 1.1 | 1.1 |
|---|---|---|---|---|---|---|---|---|
| $T$ | 0.5 | 1 | 2 | 5 | 10 | 20 | 1 | 2 |
| $p_{E03}$ | **0.336** | **0.78** | **1.40** | **2.67** | **4.04** | **6.08** | $5 \times 10^{-4}$ | 0.0049 |
| $p_{V03}$ | 0.0019 | 0.023 | 0.089 | **0.40** | **1.02** | **2.00** | $5 \times 10^{-5}$ | $3.5 \times 10^{-4}$ |
| $p_{Ma03-0}$ | $4.4 \times 10^{-5}$ | $2.6 \times 10^{-4}$ | $9 \times 10^{-4}$ | 0.006 | 0.016 | 0.033 | 0 | $1.1 \times 10^{-4}$ |
| $p_{Ma03}$ | $2.6 \times 10^{-6}$ | $2 \times 10^{-5}$ | $2 \times 10^{-5}$ | 0.001 | 0.003 | 0.008 | 0 | $6 \times 10^{-7}$ |
| $p_{Me03-0}$ | 0 | 0 | 0 | 0.002 | 0.005 | 0.014 | 0 | 0 |
| $p_{Me03}$ | 0 | 0 | 0 | $9 \times 10^{-4}$ | 0.002 | 0.003 | 0 | 0 |
| $p_{Sun}$ | 0 | 0 | 0 | 0.004 | 0.008 | 0.024 | 0 | 0 |
| $p_{ej}$ | 0 | 0 | 0 | 0 | 0 | 0 | 0 | 0 |
| $p_{V03}/p_{E03}$ | 0.0057 | 0.030 | 0.064 | 0.15 | 0.24 | 0.33 | 0.1 | 0.072 |
| $p_{Ma03-0}/p_{E03}$ | $1.3 \times 10^{-4}$ | $3.3 \times 10^{-4}$ | $6 \times 10^{-4}$ | 0.002 | 0.004 | 0.0054 | 0 | 0.023 |
| $p_{Ma03}/p_{E03}$ | $7.7 \times 10^{-6}$ | $2.5 \times 10^{-5}$ | $2 \times 10^{-5}$ | $4 \times 10^{-4}$ | 0.0008 | 0.0013 | 0 | $1.2 \times 10^{-4}$ |
| $p_{Me03-0}/p_{E03}$ | 0 | 0 | 0 | $8 \times 10^{-4}$ | 0.0012 | 0.0023 | 0 | 0 |
| $p_{Me03}/p_{E03}$ | 0 | 0 | 0 | $3 \times 10^{-4}$ | 0.0004 | 0.0005 | 0 | 0 |
| $p_{J}/p_{E03}$ | 0 | 0 | 0 | 0 | 0 | 0 | 0 | 0 |
| $p_{Sun}/p_{E03}$ | 0 | 0 | 0 | 0.001 | 0.002 | 0.004 | 0 | 0 |
| $p_{E03}/p_{M03}$ | 44.1 | 49.9 | 48.7 | 45.7 | 42.3 | 38.4 | 20 | 23.4 |
| $p_{M}/p_{M03}$ | 2.41 | 2.79 | 2.78 | 2.71 | 2.66 | 2.37 | 2.3 | 2.67 |
| $a_{0min}$, AU | 1.1 | 1.1 | 1.1 | 1.1 | 1.3 | 1.3 | 1.3 | 1.3 |
| $T$ | 5 | 10 | 20 | 50 | 1 | 2 | 5 | 10 |
| $p_{E03}$ | 0.028 | 0.050 | **0.135** | **0.210** | 0 | 0 | 0.0012 | 0.0072 |
| $p_{V03}$ | 0.0029 | 0.012 | 0.047 | 0.099 | 0 | 0 | 0 | 0.0005 |
| $p_{Ma03-0}$ | $7.7 \times 10^{-4}$ | 0.002 | 0.006 | 0.010 | 0.014 | 0.017 | 0.030 | 0.050 |
| $p_{Ma03}$ | $5.3 \times 10^{-5}$ | $1.7 \times 10^{-4}$ | $8 \times 10^{-4}$ | 0.002 | 0.009 | 0.013 | 0.025 | 0.042 |
| $p_{Me03-0}$ | 0 | 0 | $2 \times 10^{-4}$ | 0.001 | 0 | 0 | 0 | $3 \times 10^{-5}$ |
| $p_{Me03}$ | 0 | 0 | $2 \times 10^{-5}$ | $2 \times 10^{-4}$ | 0 | 0 | 0 | $2 \times 10^{-5}$ |
| $p_{Sun}$ | 0 | 0 | 0.004 | 0.048 | 0 | 0 | 0 | 0 |
| $p_{ej}$ | 0 | 0 | 0 | 0 | 0 | 0 | 0 | 0 |
| $p_{V03}/p_{E03}$ | 0.103 | 0.23 | 0.347 | **0.47** | Inf | Inf | 0 | 0.073 |
| $p_{Ma03-0}/p_{E03}$ | 0.028 | 0.042 | 0.044 | 0.048 | Inf | Inf | 25 | 7.0 |
| $p_{Ma03}/p_{E03}$ | 0.0019 | 0.0034 | 0.006 | 0.008 | Inf | Inf | 20.5 | 5.8 |
| $p_{Me03-0}/p_{E03}$ | 0 | 0 | 0.0016 | 0.006 | Inf | Inf | 0 | 0.004 |
| $p_{Me03}/p_{E03}$ | 0 | 0 | $2 \times 10^{-4}$ | 0.001 | Inf | Inf | 0 | 0.003 |
| $p_{J}/p_{E03}$ | 0 | 0 | 0 | 0 | Inf | Inf | 0 | 0 |
| $p_{Sun}/p_{E03}$ | 0 | 0 | 0.03 | 0.023 | Inf | Inf | 0 | 0 |
| $p_{E03}/p_{M03}$ | 22.8 | 19.6 | 16.4 | 18.1 | Inf | Inf | 19.1 | 21.1 |
| $p_{M}/p_{M03}$ | 2.62 | 2.49 | 2.3 | 2.3 | Inf | Inf | 2.3 | 2.3 |

**Table 3.** (Contd.)

| $a_{0min}$, AU | 1.3 | 1.3 | 1.5 | 1.5 | 1.5 | 1.5 | 1.5 | 1.5 |
|---|---|---|---|---|---|---|---|---|
| $T$ | 20 | 50 | 1 | 2 | 5 | 10 | 20 | 50 |
| $p_{E03}$ | 0.016 | 0.040 | 0 | $5 \times 10^{-4}$ | 0.0029 | 0.0094 | 0.024 | 0.047 |
| $p_{V03}$ | 0.0030 | 0.0057 | 0 | 0 | 0 | $1 \times 10^{-7}$ | $9 \times 10^{-4}$ | 0.0012 |
| $p_{Ma03-0}$ | 0.063 | 0.081 | 0.0017 | 0.0024 | 0.0057 | 0.012 | 0.021 | 0.038 |
| $p_{Ma03}$ | 0.051 | 0.062 | 0.0006 | 0.0009 | 0.0025 | 0.005 | 0.009 | 0.018 |
| $p_{Me03-0}$ | $5 \times 10^{-5}$ | $1.5 \times 10^{-4}$ | 0 | 0 | 0 | $3 \times 10^{-8}$ | $3.5 \times 10^{-8}$ | $2 \times 10^{-6}$ |
| $p_{Me03}$ | $9 \times 10^{-5}$ | $9.0 \times 10^{-5}$ | 0 | 0 | 0 | $3 \times 10^{-8}$ | $3 \times 10^{-8}$ | $3 \times 10^{-8}$ |
| $p_{Sun}$ | 0 | 0.032 | 0 | 0 | 0 | 0.004 | 0.004 | 0.032 |
| $p_{ej}$ | 0 | 0 | 0 | 0 | 0 | 0 | 0.004 | 0.008 |
| $p_{V03}/p_{E03}$ | 0.18 | 0.14 | Inf | 0 | 0 | $1 \times 10^{-5}$ | 0.046 | 0.026 |
| $p_{Ma03-0}/p_{E03}$ | 3.84 | 2.02 | Inf | 4.6 | 1.98 | 1.25 | 0.87 | 0.81 |
| $p_{Ma03}/p_{E03}$ | 3.11 | 1.53 | Inf | 1.7 | 0.88 | 0.54 | 0.39 | 0.38 |
| $p_{Me03-0}/p_{E03}$ | 0.003 | 0.0037 | Inf | 0 | 0 | $3 \times 10^{-6}$ | $1.5 \times 10^{-6}$ | $5 \times 10^{-5}$ |
| $p_{Me03}/p_{E03}$ | 0.005 | 0.0022 | Inf | 0 | 0 | $3 \times 10^{-6}$ | $1.3 \times 10^{-6}$ | $6 \times 10^{-7}$ |
| $p_J/p_{E03}$ | 0 | 0 | Inf | 0 | 0 | 0 | 0.035 | 0.009 |
| $p_{Sun}/p_{E03}$ | 0 | 0.8 | Inf | 0 | 0 | 0.43 | 0.17 | 0.68 |
| $p_{E03}/p_{M03}$ | 17.9 | 17.6 | Inf | 17.2 | 20.8 | 22.7 | 19.0 | 18.4 |
| $p_M/p_{M03}$ | 2.3 | 2.3 | Inf | 2.3 | 2.3 | 2.3 | 2.3 | 2.3 |

Designations: $p_{E03}$, $p_{V03}$, $p_{Ma03}$, $p_{Me03}$, and $p_{M03}$ are the probabilities of collisions of a planetesimal with embryos of the terrestrial planets—Earth, Venus, Mars, and Mercury—and the Moon, the masses of which are 0.3 of the present masses of the planets and the Moon, respectively. $p_{Ma03-0}$ and $p_{Me03-0}$ are the probabilities of collisions of a planetesimal with the embryos of Mars and Mercury, respectively, with such masses for the case when the orbital eccentricities of these embryos were zero. All probabilities were calculated for the time interval $T$. The other notations are the same as those in Table 2.

of initial orbits ranging from 0.7 to 0.9 AU (the total mass of these planetesimals could be not less than $0.4m_E$), the MeS$_{01}$ calculation series yielded the values $p_{V01} = 0.2$, 0.4, and 0.8 for $T = 1$, 2, and 5 Myr, respectively. These estimates suggest that the masses of the embryos of the Earth and Venus could grow twofold for 1 Myr, i.e., the mass of the Earth's embryo could increase from $0.1m_E$ to $0.2m_E$. For a zone at a distance of 0.5–0.7 AU from the Sun, the value of $p_{V01}$ was 0.36 and 1 for $T = 5$ and 20 Myr, respectively. For the MeN$_{03}$ calculation series and a zone at 0.9–1.1 AU from the Sun, the collision probability of a planetesimal with the Earth's embryo with a mass of $0.3m_E$ was $p_{E03} = 0.8$ and 1.4 for $T = 1$ and 2 Myr, respectively. For a zone at a distance of 0.7–0.9 AU from the Sun, the MeN$_{03}$ calculation series yielded the probability of a collision of a planetesimal with the Venusian embryo with a mass of 0.3 of the present Venusian mass $p_{V03} = 0.2$, 0.4–0.5, and 1–1.6 for $T = 1$, 2, and 5 Myr, respectively (the scattering in the $p_{V03}$ values is specified for the calculations with analogous initial data which differ only by an integration step and orientations of the initial orbits of planetesimals). For a zone at a distance of 0.5–0.7 AU from the Sun, the $p_{V03}$ values were equal to 0.4, 0.6, 1, and 2 for $T = 1$, 2, 5, and 20 Myr, respectively.

These estimates show that the embryos of Venus and the Earth were growing at nearly the same rate (though the Earth's embryo was growing slightly faster) and could accumulate most planetesimals with initial values of the semi-major axes of orbits ranging from 0.7 to 1.1 AU for less than 5 Myr; and Venus could accumulate most planetesimals with semi-major axes of the initial orbits ranging from 0.5 to 0.7 AU for a period not longer than 5–10 Myr. If $T \leq 5$ Myr, the ratio of the probabilities of collisions of planetesimals with the embryos of the Earth and Venus, the masses of which are 10 times less than the masses of the present planets, $p_{V01}/p_{E01}$ did not exceed 0.01 for planetesimals initially located at a distance of 0.9 to 1.1 AU from the Sun and exceeded 35 for planetesimals from the zone of 0.7–0.9 AU. Analogous data for the other zones show that, if the masses of embryos of the Earth and Venus were roughly $0.1m_E$, these embryos mostly accumulated the material from the zones initially located at 0.9–1.1 and 0.5–0.9 AU from the Sun, respectively. The material ejection in collisions of the bodies with the planets, which is ignored in the model, may

**Table 4.** Probabilities of collisions of a planetesimal with planets, the Moon, and the Sun for the time interval $T$ (Myr) for the disks composed of 250 primordial planetesimals with the semi-major axes of orbits ranging from $a_{0min}$ to $(a_{0min} + d_a)$ AU, the eccentricities $e_0 = 0.05$, and the inclinations $i_0 = e_0/2$ rad. The value $d_a = 0.2$ AU is assumed in all variants of the MeN calculation series except that with $a_{0min} = 1.5$ AU, where $d_a = 0.5$ AU is assumed. **The masses and orbital elements of the planets are equal to their present values**

| $a_{0min}$, AU | 0.3 | 0.3 | 0.3 | 0.3 | 0.5 | 0.5 | 0.5* | 0.5* |
|---|---|---|---|---|---|---|---|---|
| $e_0$ | 0.05 | 0.05 | 0.3 | 0.3 | 0.05 | 0.05 | 0.05 | 0.05 |
| $T$ | 5 | 20 | 5 | 20 | 5 | 20 | 5 | 20 |
| $p_E$ | 0.0044 | 0.008 | $5.2 \times 10^{-5}$ | 0.0032 | **0.22** | **0.31** | 0.084 | **0.10** |
| $p_V$ | **0.198** | **0.30** | 0.071 | **0.25** | **1.26** | **1.70** | **0.54** | **0.62** |
| $p_{Ma0}$ | 0.0002 | 0.0003 | $1.5 \times 10^{-6}$ | $4.5 \times 10^{-5}$ | 0.0037 | 0.0085 | 0.001 | 0.0017 |
| $p_{Ma}$ | Abs | 0.0007 | Abs | $1.7 \times 10^{-5}$ | Abs | 0.0005 | 0.003 | 0.0005 |
| $p_{Me0}$ | Abs | **0.62** | Abs | 0.136 | Abs | 0.13 | 0.038 | 0.0052 |
| $p_{Me}$ | **0.478** | **0.50** | 0.060 | 0.087 | 0.0237 | 0.053 | 0.014 | 0.022 |
| $p_{Sun}$ | **0.124** | **0.54** | **0.144** | **0.53** | 0.008 | **0.24** | 0.052 | **0.36** |
| $p_{ej}$ | 0.012 | 0.032 | 0.004 | 0.016 | 0 | 0 | 0.008 | 0.024 |
| $p_V/p_E$ | 44.9 | 38.5 | 134.7 | 78.1 | 5.66 | 5.50 | 6.39 | 6.2 |
| $p_{Ma0}/p_E$ | 0.045 | 0.043 | 0.003 | 0.014 | 0.016 | 0.027 | 0.012 | 0.017 |
| $p_{Ma}/p_E$ | Abs | 0.009 | Abs | 0.0054 | Abs | 0.016 | 0.004 | 0.005 |
| $p_{Me0}/p_E$ | Abs | 80.8 | Abs | 43.0 | Abs | 0.42 | 0.46 | 0.52 |
| $p_{Me}/p_E$ | 108.6 | 65.1 | 94.1 | 27.5 | **0.106** | 0.17 | 0.17 | 0.22 |
| $p_J/p_E$ | 0 | 4.3 | 0 | 0.007 | 0 | 0 | 0.0002 | 0.0003 |
| $p_S/p_E$ | 0 | 2.4 | 0 | 0.004 | 0 | 0 | $7.7 \times 10^{-5}$ | 0.0001 |
| $p_{Sun}/p_E$ | 28.2 | 70.8 | 2769 | 167 | 0.036 | 0.78 | 0.62 | 3.6 |
| $p_E/p_M$ | 19.7 | 18.7 | 21.4 | 18.5 | 32.9 | 28.6 | 26.5 | 22.8 |
| $p_E/p_{E01}$ | 6.51 | 6.24 | 6.62 | 5.99 | 8.26 | 7.77 | 7.55 | 6.91 |
| $p_M/p_{M01}$ | 4.37 | 4.38 | 4.33 | 4.66 | 4.86 | 4.72 | 4.89 | 4.72 |

| $a_{0min}$, AU | 0.5 | 0.5 | 0.5* | 0.5* | 0.7 | 0.7 | 0.7 | 0.7 |
|---|---|---|---|---|---|---|---|---|
| $e_0$ | 0.3 | 0.3 | 0.3 | 0.3 | 0.05 | 0.05 | 0.3 | 0.3 |
| $T$ | 5 | 20 | 5 | 20 | 5 | 20 | 5 | 20 |
| $p_E$ | 0.065 | 0.094 | **0.108** | **0.234** | **0.645** | **0.978** | **0.36** | **0.60** |
| $p_V$ | **0.572** | **0.784** | **0.992** | **1.397** | **1.22** | **2.05** | **0.52** | **1.22** |
| $p_{Ma0}$ | 0.029 | 0.0027 | 0.001 | 0.006 | 0.0097 | 0.023 | 0.005 | 0.015 |
| $p_{Ma}$ | 0.0004 | 0.001 | 0.004 | 0.002 | 0.023 | 0.0075 | 0.0016 | 0.0046 |
| $p_{Me0}$ | 0.039 | 0.12 | 0.114 | 0.171 | 0.015 | **0.066** | 0.026 | **0.096** |
| $p_{Me}$ | **0.028** | **0.049** | **0.033** | **0.065** | 0.0031 | 0.022 | 0.0067 | **0.044** |
| $p_{Sun}$ | 0.048 | **0.36** | **0.152** | **0.512** | 0.024 | **0.228** | 0.032 | **0.256** |
| $p_{ej}$ | 0.012 | 0.056 | 0.008 | 0.024 | 0.02 | 0.064 | 0.004 | 0.032 |
| $p_V/p_E$ | 8.78 | 8.34 | 9.22 | 5.68 | 1.90 | 2.09 | **1.54** | **1.22** |
| $p_{Ma0}/p_E$ | 0.020 | 0.029 | 0.010 | 0.026 | 0.015 | 0.024 | 0.014 | 0.025 |
| $p_{Ma}/p_E$ | 0.006 | 0.01 | 0.003 | 0.009 | 0.0036 | 0.0076 | 0.0045 | 0.0076 |
| $p_{Me0}/p_E$ | 0.59 | 1.32 | 1.06 | 0.730 | 0.001 | **0.068** | **0.072** | 0.16 |
| $p_{Me}/p_E$ | 0.43 | 0.52 | 0.31 | 0.278 | 0.0048 | 0.023 | 0.019 | **0.073** |
| $p_J/p_E$ | 0.0003 | 0.002 | 0 | 0.0001 | 0.0008 | 0.0047 | 0.0001 | 0.0003 |
| $p_S/p_E$ | 0.0018 | 0.0015 | 0 | $5 \times 10^{-6}$ | $3.0 \times 10^{-5}$ | $4 \times 10^{-5}$ | $5 \times 10^{-5}$ | $4 \times 10^{-5}$ |
| $p_{Sun}/p_E$ | 0.738 | 3.8 | 1.4 | 2.2 | 0.037 | 0.23 | 0.089 | 0.43 |
| $p_E/p_M$ | 21.8 | 19.4 | 24.6 | 24.0 | 32.7 | 29.6 | 26.2 | 25.7 |
| $p_E/p_{E01}$ | 6.73 | 6.24 | 7.34 | 7.19 | 8.21 | 7.91 | 7.46 | 7.42 |
| $p_M/p_{M01}$ | 4.67 | 4.53 | 4.82 | 4.96 | 4.82 | 4.77 | 4.62 | 4.63 |

**Table 4.** (Contd.)

| $a_{0min}$, AU | 0.9 | 0.9 | 0.9 | 0.9 | 0.9 | 0.9* | 0.9* | 0.9* |
|---|---|---|---|---|---|---|---|---|
| $e_0$ | 0.05 | 0.05 | 005 | 0.05 | 0.05 | 0.05 | 0.05 | 0.05 |
| $T$ | 0.5 | 1 | 2 | 5 | 20 | 1 | 5 | 20 |
| $p_E$ | **0.465** | **0.838** | **1.32** | **2.00** | **4.806** | **0.668** | **1.78** | **4.35** |
| $p_V$ | **0.167** | **0.393** | **0.749** | **1.41** | **4.267** | **0.278** | **1.25** | **2.37** |
| $p_{Ma0}$ | 0.0019 | 0.0062 | 0.012 | 0.027 | 0.087 | 0.0043 | 0.021 | 0.055 |
| $p_{Ma}$ | 0.0002 | 0.0008 | 0.0021 | 0.0075 | 0.038 | 0.0006 | 0.052 | 0.017 |
| $p_{Me0}$ | $8 \times 10^{-6}$ | 0.0019 | 0.005 | 0.029 | **0.30** | 0.0008 | 0.045 | **0.139** |
| $p_{Me}$ | $10^{-6}$ | 0.0001 | 0.001 | 0.0076 | **0.108** | 0.0002 | 0.042 | **0.124** |
| $p_{Sun}$ | 0 | 0 | 0.004 | 0.016 | **0.228** | 0.004 | 0.04 | **0.244** |
| $p_{ej}$ | 0 | 0 | 0.004 | 0.012 | 0.056 | 0 | 0.02 | 0.052 |
| $p_V/p_E$ | 0.36 | **0.47** | **0.57** | **0.70** | **0.89** | **0.42** | **0.70** | **0.54** |
| $p_{Ma0}/p_E$ | 0.0041 | 0.0074 | 0.009 | 0.014 | 0.018 | 0.0064 | 0.012 | 0.038 |
| $p_{Ma}/p_E$ | 0.0005 | 0.001 | 0.0016 | 0.0038 | 0.025 | 0.0008 | 0.0029 | 0.013 |
| $p_{Me0}/p_E$ | $1.6 \times 10^{-5}$ | 0.0023 | 0.0038 | 0.014 | 0.008 | 0.0012 | 0.025 | 0.032 |
| $p_{Me}/p_E$ | $2 \times 10^{-6}$ | 0.0003 | 0.0008 | 0.0038 | 0.062 | 0.0003 | 0.023 | 0.030 |
| $p_J/p_E$ | 0 | $1 \times 10^{-5}$ | $1 \times 10^{-5}$ | 0.001 | 0.003 | 0 | 0.0004 | 0.0005 |
| $p_S/p_E$ | 0 | $4 \times 10^{-5}$ | $4 \times 10^{-5}$ | $4 \times 10^{-5}$ | 0.0003 | 0 | $3 \times 10^{-5}$ | 0.0001 |
| $p_{Sun}/p_E$ | 0 | 0 | 0.003 | 0.008 | 0.047 | 0.006 | 0.022 | 0.056 |
| $p_E/p_M$ | 39.2 | 38.3 | 37.0 | 35.6 | 35.0 | 38.5 | 36.2 | 43.1 |
| $p_E/p_{E01}$ | 9.12 | 9.03 | 8.98 | 8.86 | 8.98 | 9.04 | 8.96 | 8.75 |
| $p_M/p_{M01}$ | 5.09 | 5.02 | 5.0 | 4.93 | 4.90 | 4.89 | 4.86 | 5.42 |

| $a_{0min}$, AU | 0.9 | 0.9 | 0.9* | 0.9* | 1.1 | 1.1 | 1.1 | 1.1 |
|---|---|---|---|---|---|---|---|---|
| $e_0$ | 0.3 | 0.3 | 0.3 | 0.3 | 0.05 | 0.05 | 0.3 | 0.3 |
| $T$ | 5 | 20 | 5 | 20 | 5 | 20 | 5 | 20 |
| $p_E$ | **0.488** | **0.579** | **0.205** | **0.748** | **0.193** | **0.56** | 0.058 | 0.090 |
| $p_V$ | **0.45** | **0.62** | **0.230** | **0.928** | **0.104** | **0.49** | 0.037 | 0.092 |
| $p_{Ma0}$ | 0.019 | 0.025 | 0.010 | 0.039 | 0.016 | 0.039 | 0.0083 | 0.011 |
| $p_{Ma}$ | 0.0057 | 0.008 | 0.0029 | 0.013 | 0.003 | 0.010 | 0.0026 | 0.0035 |
| $p_{Me0}$ | 0.015 | 0.046 | 0.0021 | 0.039 | 0.0013 | 0.03 | 0.0007 | 0.0079 |
| $p_{Me}$ | 0.0053 | 0.022 | 0.0013 | 0.056 | $3 \times 10^{-4}$ | 0.012 | 0.0002 | 0.0031 |
| $p_{Sun}$ | 0.044 | **0.236** | 0.072 | **0.28** | 0 | **0.132** | 0.052 | **0.26** |
| $p_{ej}$ | 0.012 | 0.064 | 0.012 | 0.064 | 0 | 0.032 | 0.02 | 0.064 |
| $p_V/p_E$ | **0.92** | **1.07** | **1.12** | **1.24** | **0.54** | **0.87** | 0.64 | **1.02** |
| $p_{Ma0}/p_E$ | 0.039 | 0.043 | 0.049 | **0.052** | **0.084** | **0.069** | **0.143** | **0.12** |
| $p_{Ma}/p_E$ | 0.012 | 0.014 | 0.014 | 0.017 | 0.016 | 0.018 | 0.044 | 0.039 |
| $p_{Me0}/p_E$ | **0.031** | **0.080** | 0.010 | **0.052** | 0.0065 | **0.053** | 0.012 | 0.0088 |
| $p_{Me}/p_E$ | 0.011 | **0.037** | 0.0065 | **0.074** | 0.0013 | 0.021 | 0.0033 | **0.034** |
| $p_J/p_E$ | 0.0023 | 0.004 | 0.003 | 0.011 | 0 | 0.003 | 0.0022 | 0.0040 |
| $p_S/p_E$ | $4 \times 10^{-5}$ | 0.0001 | 0.0001 | 0.0002 | 0 | $2 \times 10^{-5}$ | 0.0006 | 0.0006 |
| $p_{Sun}/p_E$ | 0.09 | 0.41 | 0.35 | 0.37 | 0 | 0.23 | 0.90 | 2.9 |
| $p_E/p_M$ | 29.0 | 26.6 | 16.3 | 27.5 | 27.5 | 26.0 | 19.5 | 16.9 |
| $p_E/p_{E01}$ | 6.47 | 5.44 | 7.49 | 7.81 | 7.44 | 7.30 | 6.30 | 5.72 |
| $p_M/p_{M01}$ | 3.63 | 4.70 | 4.76 | 4.76 | 4.56 | 4.59 | 4.53 | 4.46 |

**Table 4.** (Contd.)

| $a_{0min}$, AU | 1.1 | 1.3 | 1.3 | 1.3* | 1.3* | 1.3 | 1.3 | 1.3* |
|---|---|---|---|---|---|---|---|---|
| $e_0$ | 0.3 | 0.05 | 0.05 | 0.05 | 0.05 | 0.3 | 0.3 | 0.3 |
| $T$ | 50 | 5 | 20 | 5 | 20 | 5 | 20 | 5 |
| $p_E$ | **0.101** | **0.468** | **0.562** | **0.21** | **0.43** | 0.099 | **0.257** | **0.135** |
| $p_V$ | 0.121 | 0.226 | 0.335 | 0.087 | 0.27 | 0.038 | 0.201 | 0.051 |
| $p_{Ma0}$ | 0.012 | 0.077 | 0.086 | 0.049 | 0.078 | 0.0072 | 0.019 | 0.0094 |
| $p_{Ma}$ | 0.004 | 0.013 | 0.047 | 0.036 | 0.053 | 0.0024 | 0.0065 | 0.0034 |
| $p_{Me0}$ | 0.015 | 0.004 | 0.014 | 0.0007 | 0.015 | 0.0012 | 0.012 | 0.0012 |
| $p_{Me}$ | 0.0065 | 0.0008 | 0.0055 | 0.0019 | 0.0045 | 0.0004 | 0.005 | 0.00045 |
| $p_{Sun}$ | **0.536** | 0.02 | **0.20** | 0.052 | **0.264** | 0.08 | **0.32** | 0.092 |
| $p_{ej}$ | 0.088 | 0.008 | 0.048 | 0.024 | 0.068 | 0.012 | 0.04 | 0.016 |
| $p_V/p_E$ | **1.20** | **0.483** | **0.596** | **0.41** | **0.62** | 0.38 | **0.78** | 0.38 |
| $p_{Ma0}/p_E$ | **0.117** | **0.164** | **0.153** | 0.23 | **0.18** | 0.072 | **0.075** | **0.070** |
| $p_{Ma}/p_E$ | 0.040 | 0.029 | **0.084** | **0.17** | **0.12** | 0.024 | 0.025 | 0.025 |
| $p_{Me0}/p_E$ | 0.143 | **0.086** | 0.024 | 0.0032 | **0.033** | 0.012 | **0.046** | 0.0088 |
| $p_{Me}/p_E$ | **0.064** | 0.0017 | 0.010 | 0.0089 | 0.010 | 0.0044 | 0.020 | 0.0034 |
| $p_J/p_E$ | 0.0036 | 0.0007 | 0.0018 | 0.0007 | 0.018 | 0.0033 | 0.005 | 0.0062 |
| $p_S/p_E$ | 0.0005 | 0.0004 | 0.0004 | 0.0001 | 0.003 | 0.0004 | 0.0002 | 0.0005 |
| $p_{Sun}/p_E$ | 5.3 | 0.043 | 0.36 | 0.04 | 0.61 | 0.81 | 1.245 | 0.68 |
| $p_E/p_M$ | 16.1 | 29.3 | 26.6 | 26.0 | 26.35 | 22.2 | 22.5 | 23.40 |
| $p_E/p_{E01}$ | 5.23 | 7.70 | 7.36 | 7.29 | 7.42 | 6.76 | 6.85 | 7.04 |
| $p_M/p_{M01}$ | 4.46 | 4.82 | 4.75 | 4.69 | 4.68 | 4.59 | 4.59 | 4.58 |

| $a_{0min}$, AU | 1.3* | 1.5 | 1.5 | 1.5 | 1.5 | 1.5 | 1.5 |
|---|---|---|---|---|---|---|---|
| $e_0$ | 0.3 | 0.05 | 0.05 | 0.05 | 0.3 | 0.3 | 0.3 |
| $T$ | 20 | 5 | 20 | 50 | 5 | 20 | 50 |
| $p_E$ | **0.195** | 0.0113 | 0.031 | 0.052 | 0.0082 | 0.023 | 0.043 |
| $p_V$ | **0.112** | 0.0008 | 0.017 | 0.037 | 0.0013 | 0.0166 | 0.041 |
| $p_{Ma0}$ | 0.014 | 0.0061 | 0.011 | 0.020 | 0.0047 | 0.011 | 0.017 |
| $p_{Ma}$ | 0.0056 | Abs | 0.0092 | 0.0099 | Abs | 0.009 | 0.014 |
| $p_{Me0}$ | 0.006 | Abs | 0.0016 | 0.0026 | Abs | 0.0037 | 0.0075 |
| $p_{Me}$ | 0.003 | $4 \times 10^{-5}$ | 0.0011 | 0.0023 | $2 \times 10^{-5}$ | 0.0012 | 0.0032 |
| $p_{Sun}$ | **0.344** | 0.016 | **0.108** | **0.30** | 0.02 | **0.204** | **0.428** |
| $p_{ej}$ | 0.072 | 0 | 0.024 | 0.064 | 0.008 | 0.052 | 0.072 |
| $p_V/p_E$ | **0.576** | 0.068 | **0.549** | **0.72** | 0.154 | **0.72** | **0.96** |
| $p_{Ma0}/p_E$ | **0.074** | 0.539 | 0.38 | 0.39 | 0.568 | 0.48 | 0.40 |
| $p_{Ma}/p_E$ | 0.029 | Abs | **0.20** | **0.19** | Abs | 0.40 | 0.31 |
| $p_{Me0}/p_E$ | **0.031** | Abs | **0.052** | **0.049** | Abs | 0.163 | 0.175 |
| $p_{Me}/p_E$ | 0.015 | 0.0039 | 0.036 | 0.045 | 0.0022 | **0.052** | **0.075** |
| $p_J/p_E$ | 0.013 | 0 | 0.0043 | 0.008 | 0.036 | 0.081 | 0.05 |
| $p_S/p_E$ | 0.0077 | 0 | 0.0006 | 0.0007 | 0.0013 | 0.0024 | 0.0014 |
| $p_{Sun}/p_E$ | 1.76 | 1.42 | 3.48 | 5.77 | 0.98 | 8.9 | 9.95 |
| $p_E/p_M$ | 21.25 | 17.2 | 16.8 | 16.9 | 18.7 | 17.4 | 16.6 |
| $p_E/p_{E01}$ | 6.64 | 5.84 | 5.68 | 5.69 | 6.06 | 5.77 | 5.60 |
| $p_M/p_{M01}$ | 4.54 | 4.30 | 4.47 | 4.53 | 4.62 | 4.56 | 4.53 |

Designations: $p_E$, $p_V$, $p_{Ma}$, $p_{Me}$, $p_J$, $p_S$, and $p_M$ are the probabilities of collisions of a planetesimal with the Earth, Venus, Mars, Mercury, Jupiter, Saturn, and the Moon with the present masses, respectively. $p_{Ma0}$ and $p_{Me0}$ are the probabilities of collisions of a planetesimal with Mars and Mercury, respectively, if their orbital eccentricities were zero. All probabilities were calculated for the time interval $T$. The mark "Abs" means that this value was not calculated. The other notations are the same as those in Table 2.

slightly enhance the estimates of the time required for accumulation of the planets.

For the embryos of the Earth and Venus, the masses of which were 0.3 of the present masses of the planets, and the planetesimals with initial values of semi-major axes ranging from 0.7 to 0.9 AU, the ratio of the probabilities of their collisions $p_{V03}/p_{E03}$ was approximately 3–4, 2, 1.2–1.4, and 1 for $T = 1, 2, 5$, and 10 Myr, respectively. For planetesimals with the initial values of semi-major axes ranging from 0.9 to 1.2 AU, the ratio $p_{V03}/p_{E03}$ was approximately 0.03, 0.06, 0.15, and 0.24 for $T = 1, 2, 5$, and 10 Myr, respectively. Thus, the deepest layers of the Earth and Venus were mainly formed by accumulation of the material from the vicinity of the planet, but for the masses of these embryos about 0.3 of the final masses of the planets, up to several tens of percent of the material in low to these two embryos could come from the same zone (especially from the zone at a distance of 0.7–0.9 AU from the Sun).

For 1 Myr the embryo of Mercury with a mass of approximately 0.1 of the Mercurian mass could accumulate 2–3% of planetesimals, the initial values of semi-major axes of the orbits of which were between 0.3 and 0.5 AU. However, for the next 4 Myr, the increase in mass was smaller than that for the first 1 Myr. In the $MeN_{03}$ calculation series, for a zone between 0.3 and 0.5 AU, the $p_{Me03}$ value was around 0.03–0.05 and the values for $T = 1$ and 20 Myr differed only by 1.3 times. For the bulk mass of planetesimals in this zone $M_b = 0.1 m_E$, the product $M_b p_{Me01} = 0.003 m_E$, which is almost 20 times smaller than the mass of Mercury ($0.055 m_E$). It is larger for larger values of $M_b$. The embryo of Mercury with a mass of approximately 0.3 of the Mercurian mass also accu-mulated the planetesimals, the semi-major axes of ini-tial orbits of which were in the interval from 0.5 to 0.7 AU; and, if the initial orbit of the embryo was circular, this contribution could reach up to half the present contribution from a zone located at a distance of 0.3–0.5 AU from the Sun. For a model of the embryo growth, to provide a more rapid growth of Mercury, one may consider, along with the change in the initial distribution of planetesimals, the initial embryo of Mercury that was formed under contraction of a rarefied condensation with a mass not less than $0.02 m_E$.

According to Zharkov (2003), the bulk mass of planetesimals in the feeding zone of Mars was initially 20 times larger than the Martian mass, but most of these planetesimals were swept out from this zone by the planetesimals penetrating from the feeding zone of Jupiter and Saturn. If we assume that, in 1 Myr following the start of formation of planets, Jupiter built up a large mass and began to transfer the neighboring planetesimals to orbits intersecting the feeding zone of Mars, while the formation time for the Martian embryo was substantially smaller, we may use the data of Tables 2 and 3 to estimate the growth of the Martian embryo for 1 Myr. For a planetesimal which initially was at a distance of 1.3–1.5 AU from the Sun, the probability of its collision with the embryo of Mars with a mass of 0.1 and 0.3 of the Martian mass for a period of 1 Myr was determined as 0.004 and 0.01, respectively. For a zone with $a_{0min} = 1.5$ AU, these probabilities were several times smaller than those for $a_{0min} = 1.3$ AU and did not exceed 0.002. As has been noted above, the ratio of the numbers of planetesimals located in zones of 1.3–1.5 and 1.5–2.0 AU is approximately 1:3, if the number of planetesimals with the semi-major axes $a$ was proportional to $a^{1/2}$. Because of this, the growth of the Martian embryo with the initial mass not exceeding $0.03 m_E$ could be smaller than $0.01 m_E$ for 1 Myr, even if the total mass of planetesimals at a distance of 1.3 to 2.0 AU from the Sun was $2 m_E$.

Let us consider the growth of the embryo of Mars for a case of a relatively small mass of the material in its feeding zone (e.g., when this zone has been already swept by planetesimals from Jupiter's zone). For the $MeS_{01}$ calculation series and a zone with $a_{0min}=1.3$ AU, the $p_{Ma01}$ and $p_{Ma01-0}$ values were 0.011–0.015 and 0.035–0.053 for $T = 5$ and 20 Myr, respectively. For these values of $T$ and the values 0.011 and 0.035 for the probability of a collision of a planetesimal with the Martian embryo with a mass equal to 0.1 of the present mass of Mars, we derive $M_b p_{Ma01} = 0.002 m_E$ and $0.007 m_E$, respectively (if the bulk mass $M_b$ of planetesimals in this zone is $0.2 m_E$. For the $MeS_{01}$ calculation series and a zone with $a_{0min}=1.5$ AU, $p_{Ma01-0} = 0.0075$ and 0.032 for $T = 5$ and 20 Myr, respectively; and the $p_{Ma01}$ values were at least 6 times smaller than the $p_{Ma01-0}$ values. Prior to the feeding zone of Mars being swept by Jupiter, the $M_b$ values for zones with $a_{0min} = 1.3$ and 1.5 AU could have exceeded $0.2 m_E$ by several times. However, even if $M_b = 0.5 m_E$ and $1.5 m_E$, the considered model will hardly yield for these two zones the mass growth of the Martian embryo by more than $0.01 m_E$ from $0.01 m_E$ for 5 Myr.

For a zone with $a_{0min} = 1.3$ AU, the $p_{Ma03}$ and $p_{Ma03-0}$ values did not exceed 0.03, 0.06, and 0.08 if $T = 5$, 20, and 50 Myr, respectively; and they were several times smaller for a zone with $a_{0min} = 1.5$ AU (they did not exceed 0.006, 0.02, and 0.04, respectively). If the bulk mass of planetesimals is $0.5 m_E$ and $1.5 m_E$ in zones with $a_{0min} = 1.3$ and 1.5 AU, respectively, the mass growth of the embryo with an initial mass of $0.03 m_E$ is not higher than $0.025 m_E$ for 5 Myr. The $p_{Ma}$ and $p_{Ma-0}$ values for a zone with $a_{0min} = 1.3$ AU did not exceed 0.04 and 0.05 if $T = 5$ and 20 Myr, respectively; and they did not exceed 0.006 and 0.01, respectively, for a zone with $a_{0min} = 1.5$ AU. In other words, these values were not substantially larger than those for the Martian embryo that was three times smaller in mass. If the bulk mass of planetesimals is $0.5 m_E$ and $1.5 m_E$ in zones with $a_{0min} = 1.3$ and 1.5 AU, respectively, we find that the Martian embryo grows in mass by not more than $0.04m_E$ for 20 Myr, even if its mass was equal to that of Mars. Because of this, even for the bulk mass of plan-etesimals in these zones equal to $2m_E$, the growth of the bulk of the Martian mass could be extended for tens of millions of years; consequently, within this model, Mars would also be growing to its present mass slower than the other terrestrial planets. Actually, it is the influence of Jupiter that could diminish the mass of planetesimals in the feeding zone of Mars by several times for a period smaller than several million years. The data on the composition of Mars suggest that the bulk of the Martian mass was formed for a time not exceeding 10−20 Myr (Elkins-Tanton, 2008; Mezzer et al., 2013; Bouvier et al., 2018) or even 5 Myr (Elkins-Tanton, 2018). Consequently, we may suppose that a relatively large embryo of Mars with a mass several times smaller than that of Mars (e.g., $≥0.03m_E$) could be formed directly as a result of contraction of a rarefied condensation.

It was possible that the initial mass of the material in the feeding zone of Mars was not smaller than that in the feeding zone of the Earth, and the Martian embryo that was formed under contraction of the parental condensation was not smaller than that of the Earth. Recent papers on the formation of rarefied condensations (see, a review by Ipatov (2017)) consider it acceptable that massive condensations can be formed. For example, Lyra et al. (2008) considered the formation of condensations with masses of $\sim(0.1−0.6)m_E$. Due to the sweeping of the feeding zone of Mars caused by the influence of Jupiter, the formation of the bulk of the Martian mass could be completed prior to the formation of the bulk of the Earth's mass; however, weak bombardment of an almost-formed Mars by planetesimals formed in its feeding zone could occur later than the bombardment of the Earth by planetesimals formed in its feeding zone.

For Mercury, the ratio of the masses of the planetary embryo formed under contraction of a condensation and the planet could also be larger than that for the Earth. However, a possible smaller absolute mass of Mercury's embryo formed under contraction and a smaller mass of the material in its feeding zone, as compared to the Earth, contributed to a smaller final mass of Mercury compared to the mass of the Earth.

## PROBABILITIES OF INFALLS OF PLANETESIMALS ONTO ALMOST-FORMED TERRESTRIAL PLANETS

The estimates presented in this section are related to the final stages of the formation of planets. They are based on the MeN calculation series, within which the influence of the gravity of the present planets was accounted for in the study of migration of planetesimals. To the moment of formation of the bulk of the planetary masses, the mean eccentricities of planetesimals could exceed 0.2. Because of this, the calcula-tions were performed for $e_0 = 0.05$ and 0.3. The evolu-tion of orbits of planetesimals from zones with $a_{0min} = 0.5$, 0.7, and 0.9 was strongly influenced by the Earth and Venus. For these values of $a_{0min}$, the probabilities of collisions of planetesimals with these planets for $e_0 = 0.05$ usually differed from those for $e_0 = 0.3$ by less than 2−3 times. Consequently, we may expect that, in this case, relatively close values of the probabilities may also be obtained for the other values of $e_0$ between 0.05 and 0.3. If the masses of planets were smaller, the material mixing could be less intensive; however, accounting for the mutual gravitational influence of planetesimals would increase the mixing of planetesimals originating from different zones. The sum $p_E + p_V$ was usually not smaller than 1 for $T ≥ 5$ Myr and $a_{0min} = 0.7$ and 0.9 AU in the both MeN and MeS_{01} calculation series. This is indicative of the fact that most planetes-imals which initially had been at a distance of 0.7−1.1 AU from the Sun fell onto the growing Earth and Venus in 5 Myr.

The ratio $p_V/p_E$ of the probabilities of collisions of planetesimals with the Earth and Venus was mainly within a range of 0.5 to 1.9 for $T ≥ 2$ Myr and $a_{0min} = 0.7−1.1$ AU. For $a_{0min} = 1.3$ AU, this ratio was also close to this range. Consequently, the material portions from different parts of this zone at 0.7−1.5 AU from the Sun, which were incorporated into the Earth and Venus at the final stages of the formation of these planets, differed not more than twofold. For the primordial planetesimals with $a_{0min} = 0.3$ and 0.5 AU, the portion of planetesimals that fell onto Venus was at least several times larger than the portion that fell onto the Earth.

The ratio of the masses of Mercury and Mars to that of the Earth are 0.055 and 0.107, respectively. The ratio $p_{Ma0}/p_E$ was in a range of 0.0535−0.214 (i.e., it differed less than twofold from the mass ratio of Mars and the Earth) in some calculations with $a_{0min} = 1.1$ or 1.3 AU, while the ratio $p_{Ma}/p_E$ was in the same range for some runs with $a_{0min} = 1.3$ or 1.5 AU. The values of $p_{Ma}$ and $p_{Ma0}$ were highest for $a_{0min} = 1.3$ AU. In other words, at the final stages of formation of the planets, planetesimals which initially had been at 1.1−2.0 AU from the Sun could be incorporated into the Earth and Mars in a ratio not very different from the mass ratio of these planets.

The ratio $p_{Me}/p_E$ was close to 100 for $a_{0min} = 0.3$ AU, while $p_{Me}$ was approximately 0.5 for $a_{0min} = 0.3$ AU, $e_0 = 0.05$, and $T = 5$ Myr and remained almost the same for $T = 20$ Myr. For $a_{0min} ≥ 0.5$ AU, the value of $p_{Me}/p_E$ was substantially smaller than that for $a_{0min} = 0.3$ AU. For $a_{0min} ≥ 0.7$ AU, in some calculations the values of $p_{Me}/p_E$ and $p_{Me0}/p_E$ were in a range of 0.027−0.11 (i.e., in a range of 0.5−2 relative to the ratio of the masses of Mercury and the Earth). How-ever, as has been noted in the previous section, prior to the formation of large planetary embryos, Mercury and Mars accumulated mainly the material from the vicinity of their orbits. The results of calculations support the hypothesis that, at the final stages of their formation, the terrestrial planets and the Moon incorporated the material which initially had been near the orbits of the other planets.

## INFALL OF PLANETESIMALS ONTO THE SUN AND THE GIANT PLANETS AND EJECTION OF PLANETESIMALS INTO HYPERBOLIC ORBITS

Some planetesimals which initially had been mainly at 0.3–0.5 AU from the Sun fell onto the Sun. In the $MeS_{01}$ calculation series, the probability of the infall of a planetesimal onto the Sun was $p_{Sun} \approx 0.03$ for $T = 5$ Myr and $a_{0min} = 0.3$ AU. For larger values of $a_{0min}$, this calculation series yielded smaller values of $p_{Sun}$.

For planetesimals with $a_{0min} = 0.3$ AU, the $MeN_{03}$ calculation series yielded $p_{Sun} = 0.04, 0.12,$ and $0.21$ for $T \geq 10$ Myr and $a_{0min} = 0.3$ AU, the probability of a colli-sion of a planetesimal with the Sun exceeded the total probability of collisions of this planetesimal with all planetary embryos. For primordial planetesimals which initially had been at larger distances from the Sun, the probability $p_{Sun}$ was lower. Specifically, for $T = 20$ Myr, the $MeN_{03}$ calculation series yielded $p_{Sun} = 0.056, 0.024, 0.004,$ and $0.004$ for $a_{0min} = 0.5, 0.9, 1.1,$ and $1.5$ AU, respectively.

For the present masses of the planets (the MeN calculation series) and $T = 20$ Myr, the values of $p_{Sun}$ exceeded 0.1 for all of the considered values of $a_{0min}$; moreover, in all except two variants of calculations, they were not smaller than 0.2. However, in some calculations with this value of $T$, the collision probabilities for planetesimals and planets exceeded 1. If the planetesimals which have collided with the planets and the Sun are removed from calculations, the result of the MeN series for the model with $e_0 = 0.05$ is that 12% of planetesimals which initially were near $a_{0min} = 0.3$ AU from the Sun should collide with the Sun in 5 Myr, while approximately 70% of the remaining planetesimals are to fall onto Mercury and Venus. Based on calculations for planetary embryos, one may expect that not less than 10% of planetesimals which initially had been closer than 0.5 AU to the Sun could fall onto the Sun. For $e_0 = 0.3$, $a_{0min} = 0.3$ AU, and $T = 20$ Myr, the value $p_{Sun} = 0.5$ exceeded the sum of probabilities of collisions of planetesimals with planets. In other words, at increase of the orbital eccentricity of planetesimals (e.g., due to the mutual gravitational influence and the influence of bodies penetrating this zone from remote regions), the probability of a collision of a planetesimal with the Sun could several times exceed 0.1. In considered variants of calculations, the embryos relatively rapidly scooped out the planetesimals which initially had been at distances of 0.3–0.5 AU from the Sun, if the embryos were close to the present planets in mass.

In the zone with $a_{0min} = 0.5$ AU and $e_0 = 0.05$, most planetesimals would have fallen onto Venus and the Earth (under an approximate ratio of 6 to 1) for roughly 5 Myr, i.e., before a noticeable portion of planetesimals would have fallen onto the Sun. For $a_{0min} = 0.5$ AU and $e_0 = 0.3$, the similar infall of most planetesimals onto Venus and the Earth would have occurred under $p_{Sun} \approx 0.1$. For $a_{0min} = 0.7$ and $0.9$ AU, the value of $p_{Sun}$ would have not exceeded 0.04 by the time when $(p_E + p_V)$ reached 1. For $a_{0min} = 1.1$ AU, $e_0 = 0.05$, and $T = 20$ Myr, the value of $p_{Sun}$ was near 0.1, while $(p_E + p_V) \approx 1$. For $a_{0min} = 1.1$ AU, $e_0 = 0.3$, and $T = 50$ Myr, only approximately 25% of the primordial planetesimals would have fallen onto the planets, roughly half of them would have collided with the Sun, and 9% of all primordial planetesimals would have been ejected into hyperbolic orbits. For $a_{0min} = 1.3$ AU, $e_0 = 0.05$, and $T = 20$ Myr, the value of $p_{Sun}$ was approximately 0.2–0.25, while $(p_E + p_V) \approx 1$. For $a_{0min} = 1.3$ AU, $e_0 = 0.3$, and $T = 20$ Myr, the value of $p_{Sun}$ was in a range of 0.3–0.35, while $(p_E + p_V)$ was approximately 0.3–0.45 (the scattering in the values for two variants of calculations), i.e., most primordial plane-tesimals would have fallen onto the planet and the Sun in a little more than 20 Myr. Summing up the above results, we may expect that a portion of planetesimals that would have fallen onto the Sun could exceed 10%, if their initial distances from the Sun are in ranges of 0.3–0.5 and 1.1–2.0 AU.

A portion of planetesimals ejected into hyperbolic orbits did not exceed 10%. The probability of a collision with Jupiter for a planetesimal which initially had been in the feeding zone of the terrestrial planets was not more than several percent of the probability of its collision with the Earth (see the lines for $p_J/p_{E01}$, $p_J/p_{E03}$, and $p_J/p_E$ in Tables 2–4), while the probabilities of collisions of planetesimals with Saturn (the line for $p_S/p_E$ in Table 4) were on average an order of magnitude smaller than those for Jupiter.

## PROBABILITIES OF COLLISIONS OF PLANETESIMALS WITH THE LUNAR EMBRYO

The studies by Ipatov (2018) support the multiimpact model developed by some specialists, according to which the embryos of the Earth and the Moon were growing under multiple collisions of planetesimals with these embryos, and the lunar embryo mainly grew in mass due to the material ejected from the Earth's embryo. As distinct from the other papers focused on this model (and discussed by Ipatov (2018)), we supposed that the embryos of the Earth and the Moon were formed due to contraction of a common rarefied condensation. In this section, we discuss the ratio of the probabilities of collisions of planetesimals with the embryos of the Earth and the Moon. Because of the weaker gravity of an embryo, the lunar embryo may lose a much larger mass in high-speed encounters than the Earth's embryo. The results of collisions of planetesimals with embryos are not consid-ered below and can be a subject of separate studies.

In comparison of the growth of two celestial bodies, the increase in the body's mass is proportional to the squared effective radius $r_{eff}$ (the circle area with a radius $r_{eff}$). The effective radius $r_{eff}$ is an impact param-eter at which a planet (a celestial body) is reached. It is calculated with the formula

$$r_{ef} = r \cdot (1 + (v_{par}/v_{rel})^2)^{1/2}, \qquad (1)$$

where $v_{par}$ is the parabolic velocity on the surface of a planet with a radius $r$, while $v_{rel}$ is the relative velocity at infinity (Okhotsimskii, 1968, pp. 36–37). If $v_{rel} > v_{par}$ (e.g., for comets infalling onto the Earth from highly eccentric orbits), $r_{ef}$ is close to $r$. In this case, the ratio of the collision probabilities for two celestial bodies is close to $(m_r/\rho_{rel})^{2/3}$, where $m_r$ and $\rho_{rel}$ are the ratios of their masses and densities, respectively. For the Earth and the Moon, $m_r/\rho_{rel} \approx 49.2$ and $(m_r/\rho_{rel})^{2/3} \approx 13.4$. If the densities of the both celestial bodies are the same, ($\rho_{rel} = 1$) and the masses differ by 10 times, $(m_r/\rho_{rel})^{2/3} \approx 10^{2/3} \approx 4.64$. For $\rho_{rel} = 1$ and the mass ratio of celestial bodies equaled to 0.3, we have $(m_r/\rho_{rel})^{2/3} \approx 3.33^{2/3} \approx 2.23$.

If the relative velocities $v_{rel}$ are small and $(v_{par}/v_{rel})^2$ is much larger than 1, $r_{ef}$ is close to $r(v_{par}/v_{rel})$, where $v_{par} = (2Gm/r)^{1/2}$ and $m$ is the mass of a planet with a radius $r$. Then, $r_{ef}$ is close to $r(v_{par}/v_{rel}) = r(2Gm/r)^{1/2}/v_{rel} = (8Gr\pi\rho/3)^{1/2} r^2/v_{rel}$, where $m = (4/3)\pi\rho r^3$ and $\rho$ is the density of a planet. In this case, with the same values of $v_r$, the ratio of the squares of $r_{eff}$ for the masses and densities of the embryos of the Earth and the Moon is $m_r^{4/3}\rho_{rel}^{-1/3} \approx 297$.

From the probabilities of all collisions of planetes-imals with the Earth and the Moon or their embryos for the time $T$, the ratio of the probabilities for the components of this satellite system was calculated. If the masses of embryos of the terrestrial planets and the Moon were 10 times smaller than their present masses (the MeS$_{01}$ calculation series), then the values of $p_{E01}/p_{M01}$ were in a range of 15–24 at $0.3 \leq a_{0min} \leq 1.1$ AU. In Table 2, only a very small value of $p_{E01}$ for $a_{0min} = 1.3$ AU and $T = 50$ Myr differed from zero. For $a_{0min} = 1.5$ AU the ratio $p_{E01}/p_{M01}$ was in the range 14−20. The highest values of this ratio, approximately 24, were obtained for $a_{0min} = 0.9$ AU, since in this case planetesimals fell onto the Earth's embryo mainly from the close orbits with smaller eccentricities (and, consequently, smaller relative velocities) than the planetesimals which had come from the periphery of the feeding zone of the ter-restrial planets. Even the highest values, close to 24, differ less than twofold from a value of 13.4, which cor-responds to the ratio of the squared radii.

In Table 2, there are also values of the ratio $p_M/p_{M01}$ of the probabilities of collisions of planetesimals with the Moon and its embryo with a mass 10 times smaller than that of the Moon and the ratio $p_E/p_{E01}$ of the probabilities of collisions of planetesimals with the Earth and its embryo with a mass 10 times smaller than that of the Earth. When calculating these probabilities, we considered the arrays of orbital elements of planetesimals and planetary embryos obtained in the MeS$_{01}$ calculation series. Since the mass of the Earth's embryo is greater than that of the lunar embryo (and, consequently, the parabolic velocity on the surface of the Earth's embryo is higher), the ratios $p_E/p_{E01}$ were larger than $p_M/p_{M01}$ (in some cases, even almost two-fold), though the corresponding masses for the both probability ratios differed by 10 times.

In the MeN$_{03}$ calculation series, the ratio $p_{E03}/p_{M03}$ was larger than the ratio $p_{E01}/p_{M01}$ in the MeS$_{01}$ calcu-lation series, though the corresponding ratios of the masses and densities of the embryos of the Earth and the Moon were the same in the both series. In the MeN$_{03}$ calculation series, the masses of the embryos were three times larger than those in the MeS$_{01}$ series. At maximum, $p_{E03}/p_{M03}$ was 54. For comparison, the highest value of $p_{E01}/p_{M01}$ was 24. These numbers characterize the increase of the relative growth of the Earth's embryo, as compared to that of the lunar embryo, due to the infall of planetesimals (the lunar embryo could also grow at the expense of the material ejected from the Earth's embryo) with the considered increase of masses of these embryos. For $a_{0min} \geq 1.1$ AU, the values of $p_{E03}/p_{M03}$ were smaller than those for $a_{0min} \leq 0.9$ AU, since the characteristic orbital eccen-tricities of planetesimals crossing the Earth's orbit are larger for planetesimals with $a_0 \geq 1.1$ AU. In other words, the ratio of the probabilities of infalls onto the embryos of the Earth and the Moon was larger for planetesimals with $a_0 \geq 1.1$ AU. The ratio $p_M/p_{M03}$ in the MeN$_{03}$ calculation series did not exceed 3 (for comparison, the ratio of the squares of the corre-sponding radii is 2.23). In this calculation series, the ratio $p_{E03}/p_{M03}$ was larger than the ratio $p_{E01}/p_{M01}$ in the MeS$_{01}$ series, though the corresponding ratios of the masses and densities of the embryos of the Earth and the Moon were the same in the both series.

When simulating the evolution of orbits of plane-tesimals at the present masses of the planets, on the basis of the arrays of the orbital elements of planetesi-mals in the course of evolution, we also calculated the values of the probability $p_{E01}$ of a collision of a plane-tesimal with an embryo of $0.1 m_E$ in mass on the Earth's orbit and the probabilities $p_{M03}$ of a collision of

**Table 5.** Characteristic relative velocities $v_{rel}$ of planetesimals entering the action sphere of the Earth and the ratios $v_{rel}/v_c$ (where $v_c$ is the velocity of the Earth moving along its heliocentric orbit) for several values of the ratio $p_E/p_{E01}$ of the probabilities of collisions of planetesimals with the Earth and its embryo with a mass 10 times smaller than that of the Earth

| $p_E/p_{E01}$ | 9.12 | 9 | 8.5 | 8 | 7.5 | 7 | 6.5 | 6 | 5.5 | 5.23 |
|---|---|---|---|---|---|---|---|---|---|---|
| $v_{rel}$, km/s | 8.65 | 8.81 | 9.55 | 10.43 | 11.51 | 12.90 | 14.78 | 17.57 | 22.46 | 27.36 |
| $v_{rel}/v_c$ | 0.290 | 0.296 | 0.32 | 0.35 | 0.39 | 0.43 | 0.50 | 0.59 | 0.75 | 0.92 |

a planetesimal with the lunar embryo with a mass of 0.3 of the lunar mass. In Table 4, the ratio $p_E/p_{E01}$ was in a range of 5.23 to 9.12 (including the ranges 5.68–9.12 and 5.23–7.81 for $e_0 = 0.05$ and 0.3, respectively). When studying the migration of planetesimals from the feeding zone of Jupiter and Saturn, Marov and Ipatov (2018) obtained that the ratio of the probabilities $p_E/p_{E01}$ was in a range from $5.5 \approx 10^{0.74}$ to $5.8 \approx 10^{0.76}$. Considering formula (1) and the ratios $p_E/p_{E01} = (r_{effE}/r_{effE01})^2$ (where $r_{effE}$ and $r_{effE01}$ are the effective radii of the Earth and its embryo with a mass of $0.1m_E$, respectively) we derive the relative velocity of a planetesimal entering the action sphere of the Earth

$v_{rel} \approx 11.19 \times (1-10^{-4/3}(p_E/p_{E01}))^{1/2}/(10^{-2/3}(p_E/p_{E01})-1)^{1/2}$

(expressed in kilometers per second). Below, in Table 5, we present the values of $v_{rel}$ and $v_{rel}/v_c$ (where $v_{rel}/v_c$ is the ratio of $v_{rel}$ to the velocity of the Earth moving along its heliocentric orbit $v_c \approx 29.78$ km/s) for several values of $p_E/p_{E01}$.

As follows from the data in Table 5, the orbital eccentricities of planetesimals that fell onto the Earth mainly exceeded 0.3. For comparison, Nesvorný et al. (2017) found in a number of models that, for asteroids which initially had semi-major axes of their orbits between 1.6 and 3.3 AU, the mean velocities of impacts on the Earth range from 21 to 23.5 km/s.

In the MeN calculation series, the ratio $p_M/p_{M03}$ of the probabilities of collisions of planetesimals with the Moon and its embryo, the mass of which was 0.3 of the lunar mass, did not exceed 5.1, i.e., it hardly differed from the ratio of the squares of the radii, equal to 4.64. The ratio $p_E/p_M$ of the probabilities, corresponding to the ratio of the masses of planetesimals that fell onto the Earth and those that collided with the Moon, varied from 16 (for $a_{0min} = 1.1$ AU, $e_0 = 0.3$, and $T = 50$ Myr) to 43 (for $a_{0min} = 0.9$ AU, $e_0 = 0.05$, and $T = 20$ Myr) in the considered variants of the MeN calculation series. This ratio $p_E/p_M$ was on average slightly smaller than the ratio $p_{E03}/p_{M03}$ for the MeN$_{03}$ calculation series. This difference is apparently caused by the larger mean eccentricities of orbits of planetesimals (as compared to those for embryos), which crossed the Earth's orbit, since the masses of planets are larger than the masses of their embryos. When studying the migration of planetesimals from the feeding zones of Jupiter and Saturn, Marov and Ipatov (2018) found that the ratio of the probabilities of collisions of a planetesimal with the Earth and the Moon was in a range between 16 and 17.

In all considered variants of calculations, the ratio of the probability of a collision of a planetesimal with the Earth's embryo (or the Earth) to that with the lunar embryo (or the Moon) was smaller (sometimes, by several times) than 81 (the ratio of the masses of the Earth and the Moon and the ratio of the masses of their embryos in the calculations). If all these collisions of planetesimals with the Earth and the Moon resulted in coagulation, the relative growth of the Moon due to these collisions would be stronger than that of the Earth. To compare the relative growth of embryos of the Earth and the Moon, the results of simulations of collisions of planetesimals with these embryos should be used. Since the mass and gravity of the lunar embryo are smaller (relative to those of the Earth), some high-speed impacts of planetesimals on the Moon could result in the ejection of material from the lunar surface and even in a decrease in its mass. To explain the iron-depleted composition of the Moon, we could suppose that its mass increased mainly at the expense of the bodies ejected from the Earth's surface under collisions of planetesimals with the Earth (Ipatov, 2018). As distinct from the megaimpact model (Hartmann and Davis, 1975; Cameron and Ward, 1976; Canup and Asphaug, 2001; Canup, 2004, 2012; Canup et al., 2013; Cuk and Stewart, 2012; Cuk et al., 2016; Barr, 2016), Ipatov (2018) considered a large number of collisions of planetesimals with the embryos of the Earth and the Moon that were formed as a result of contrac-tion of a common rarefied condensation.

As the Earth, the Moon incorporated into its com-position the same material from almost the entire feeding zone of the terrestrial planets. The iron-depleted composition of the Moon is caused by a sub-stantial contribution of the material ejected from the Earth's embryo. This ejection was to occur mainly at the stage when the iron core had been already formed in the Earth's embryo.

In the composition of the Earth and the Moon, a portion of the material which came from beyond Jupi-ter's orbit is relatively small. If the probability of a col-lision with the Earth of a planetesimal from the feed-ing zone of Jupiter and Saturn did not exceed $10^{-5}$ (Marov and Ipatov, 2018; Ipatov, 2019), the total mass of bodies which fell onto the Earth does not exceed $0.001m_E$ if the total mass of planetesimals in this zone was $100m_E$. Some decrease of this estimate may be caused by the fact that a substantial portion of these planetesimals fell onto the embryo smaller than the Earth in mass. If the probability of a collision with the Earth of a planetesimal from the feeding zone of Uranus and Neptune is $10^{-6}$ (Ipatov, 2019), then the total mass of bodies which fell onto the Earth is $0.0001 m_E$ for the total mass of planetesimals in this zone equal to $100 m_E$. The infall of planetesimals from the feeding zone of Jupiter and Saturn onto the embryos of the Earth and the Moon occurred mainly for the first millions of years of the Solar System's lifetime, when the Earth and the Moon had not yet been formed, and the material of these planetesimals could be incorporated into the inner zones of the Earth and the Moon. However, if the relative collision velocities are high and the embryos are small in mass, a substantial portion of the material of these planetesimals, could be ejected after collisions especially with the lunar embryo. A consid-erable part of the bodies from the feeding zones of Uranus and Neptune, which had collided with the Earth and the Moon, fell onto the almost-formed Earth and Moon rather than onto their small embryos.

FORMATION OF THE TERRESTRIAL PLANETS

Based on calculations within the above models and the results presented in the Introduction, we will discuss one of the possible processes of formation of the terrestrial planets. Let us consider the model of the formation of planets proposed by Safronov (1972). In addition to the latter, we will consider the formation of embryos of Uranus and Neptune near Saturn's orbit suggested by Zharkov and Kozenko (1990) and the relatively gentle migration of these embryos, under the inf luence of their interactions with planetesimals, to the present orbits of Uranus and Neptune (Ipatov, 1991a, 1991b, 1993a, 2000). In our opinion, with this model, we can explain peculiar features in the formation of the terrestrial planets. For this, it is not necessary to consider the migration of Jupiter to the Martian orbit, which took place in the Grand Tack model; and the sharp changes in Jupiter's orbit caused by the giant planets falling into a resonance, as in the Nice model, can be also avoided.

According to Chambers (2006), Jupiter's embryo with a mass of $10 m_E$ was formed in approximately 1 Myr, while the Earth's embryo with a mass of $0.1 m_E$ was formed in approximately 0.1 Myr. Having reached a mass of $10 m_E$, Jupiter's embryo could relatively quickly increase its mass by gas accretion. As we have concluded from our estimates above, in 1 Myr the masses of embryos of the Earth and Venus could grow twofold from $0.1 m_E$ and $0.08 m_E$, respectively; i.e., the mass of the Earth's embryo could reach around $0.2 m_E$. At the same time, the bodies from the feeding zone of Jupiter could start, to penetrate at perihelion into the feeding zone of the terrestrial planets. The Earth and Venus could acquire a substantial part (more than a half) of their masses in 5 Myr. In particular, during this time, most planetesimals which initially had been at a distance of 0.7−1.1 AU from the Sun, fell onto the growing Earth and Venus.

As the simulations of the migration of planetesimals from the feeding zone of Jupiter and Saturn show, the majority of planetesimals left this zone in several millions of years. In the study of the gravitational influence of Jupiter, Saturn, and the terrestrial planets, the evolution of disks of planetesimals corresponding to this zone was completed in a time less than 4 Myr (Marov and Ipatov, 2018). The calculations show that, in the model containing all planets, the orbital evolution of planetesimals may take much more time than for the runs in the absence of Uranus and Neptune (Ipatov, 2019). However, in these calculations, the main contribution to the probabilities of collisions with the embryos of the terrestrial planets for planetesimals from the feeding zones of Jupiter and Saturn also fell on the first million years after the formation of a significant mass of Jupiter (this could take place in 1−2 Myr after the beginning of the formation of the Solar System). Some planetesimals from the feeding zones of Uranus and Neptune fell onto the Earth over hundreds of millions of years and could even remain in the Solar System to the present time. When planetesimals from the feeding zone of Jupiter and Sat-urn fell onto the embryos of the terrestrial planets, these embryos had not acquired the present masses of the planets, and the material (including water and volatiles) from this zone could be accumulated in the inner layers of the terrestrial planets and the Moon.

As has been noted in the Introduction, the sweeping of the asteroid belt can easily be explained by the gravitational influence of planetesimals f lying into this belt from the feeding zones of Jupiter and Saturn and by the shift of resonances caused by the decrease of the semi-major axis of Jupiter, which ejected planetesimals into hyperbolic orbits. The large orbital eccentricities of Mercury and Mars can be explained by the gravitational influence of massive planetesimals reaching the orbits of these planets from the feeding zones of Jupiter and Saturn. A high content of iron in the core of Mercury is usually explained by the loss of most of the mass of the silicate shell under high-velocity impacts. Note that, according to the present calculations, some of the planetesimals from the vicinity of Mercury's orbit passed relatively close to the Sun before collisions with Mercury and could lose some portion of their silicate material during these passes.

Ipatov (2018) considered a model, according to which the embryos of the Earth and the Moon were formed from a common rarefied condensation with a mass larger than $0.1 m_E$. In a collision of two parental condensations, the resulting condensation acquired the angular momentum sufficient for forming a large satellite (the lunar embryo). The papers dealing with the condensation formation are reviewed by Ipatov (2017), who considered the formation of satellite systems of small bodies from rarefied condensations. The formation of satellite systems of small bodies from rarefied condensations, which acquired at collisions the angular momentum needed for such formation, was also discussed by Ipatov (2010b). The formation models con-sidered by Ipatov (2017, 2018) for the satellite systems of small bodies and the Earth–Moon system are similar.

According to Elkins-Tanton (2018), Mars grew to approximately its present size in less than 5 Myr. The estimates shown in Tables 2–4 suggest that, in our model, Mars was growing slower than the Earth and Venus, while some planetesimals in its feeding zone could also remain for 50 Myr. Consequently, we may suppose that a rather large embryo of Mars (e.g., not smaller than $0.02m_E$ in mass) may be the result of contraction of a condensation, while planetesimals from the feeding zone of Jupiter and Saturn contributed to a more rapid removal of planetesimals from the feeding zone of Mars. We may suppose that the embryo of Mercury with a mass of approximately $0.02m_E$ is also the result of contraction of a condensation.

Unlike the parental condensation of embryos of the Earth and the Moon, the parental condensation of the Martian embryo had no large angular momentum and could produce only small satellites (Phobos and Deimos). The angular momenta of parental condensations of Mercury and Venus were not sufficient even for producing small satellites. As was noted by Ipatov (2017, 2018), the angular momenta of primordial condensations, which had been formed from the protoplanetary disk, were not sufficient for producing a satellite system, while to acquire the angular momentum needed for formation of a satellite system, the conden-sation should collide with another condensation close in mass. For Mars and Mercury, the ratio of the mass of the planetary embryo resulting from the condensa-tion contraction to the mass of the planet could be larger than that for the Earth.

The portion of planetesimals which fell onto the Sun could exceed 10%, if their initial distances from the Sun ranged from 0.3 to 0.5 and from 1.1 to 2.0 AU. Less than 10% of planetesimals from the feeding zone of the terrestrial planets were ejected into hyperbolic orbits, and the ratio of the number of planetesimals that collided with Jupiter and Saturn to those that collided with the Earth did not exceed several percent.

If the planetary embryos were 10 times smaller than the present terrestrial planets in mass, these embryos accumulated planetesimals, the semi-major axes of which differed from that of the planetary embryo by less than 0.1 AU. If the masses of planetary embryos were three times smaller than those of the present terrestrial planets, the probabilities of collisions with the Earth and Venus for planetesimals which initially had been at a distance of 0.7–0.9 AU from the Sun were no more than twofold different for the time interval $T > 2$ Myr. For the other initial distances of planetesimals from the Sun, there was no such closeness of the sources of planetesimals that fell onto these embryos; however, each of the embryos could accumulate planetesimals from different regions within the feeding zone of the terrestrial planets.

Planetesimals which came from beyond Jupiter's orbit to the feeding zone of the terrestrial planets enhanced the orbital eccentricities and inclinations of planetesimals in this feeding zone. This excitation of orbits of planetesimals in the feeding zone of the terrestrial planets mainly occurred when the masses of embryos of these planets had not yet reached the current masses of the planets, since the excitation is more efficient for smaller masses of embryos. The mutual gravitational influence of planetesimals in the feeding zone of the terrestrial planets could also substantially intensify the growth of their orbital eccentricities and the material mixing in this zone as compared to the calculations presented in Tables 2–4.

## CONCLUSIONS

As distinct from the earlier modeling of the evolution of disks composed of bodies coagulating under collisions, the accumulation of the terrestrial planets was studied based on the data of the other calculations. We simulated the migration of planetesimals from the feeding zone of the terrestrial planets, which was divided into seven regions depending on the distance to the Sun. The gravitational influence of all planets was taken into account. In some variants of calculations, instead of the terrestrial planets, their embryos with masses 0.1 and 0.3 of the current masses of the planets were considered. In calculations, the planetes-imals and planets were assumed to be mass points, and their collisions were not modeled. The arrays of orbital elements of migrated planetesimals were obtained with a step of 500 yr and used in calculations of the probabilities of their collisions with planets, the Moon, and their embryos. This approach enables us to calculate more accurately the probabilities of collisions of planetesimals with planetary embryos for some evolutionary stages. When studying the composition of planetary embryos accumulated planetesimals which initially were at different distances from the Sun, we considered the narrower source zones of planetesimals, as compared to those used earlier, and studied the temporal changes in the composition of planetary embryos rather than only the final composition of planets.

The embryos of the terrestrial planets, the masses of which were approximately a tenth of the current masses of planets or less, accumulated planetesimals mainly from the vicinities of their orbits. The inner layers of a terrestrial planet were mainly formed of the material from the vicinity of its orbit. When planetesimals from the feeding zones of Jupiter and Saturn fell onto the embryos of the terrestrial planets, these embryos had not yet acquired the masses of the present planets and the material (including water and volatiles) from this zone could get into the inner layers of the terrestrial planets and influence their composition.

For the masses of embryos of the Earth and Venus of about one third of those of the present planets, the probabilities of infalls onto these embryos for planetesimals which were formed at a distance of 0.7−0.9 AU from the Sun differed less than twofold.

In the model in which the bodies coagulated with the planets in any collisions, the Earth and Venus could acquire a substantial portion (more than half) of their masses in 5 Myr. In particular, for this time, most planetesimals which were at a distance of 0.7−1.1 AU from the Sun fell onto the growing Earth and Venus. If the material ejection in collisions of the bodies with planets is accounted for, the time for accumulation of the planets may increase.

The total mass of planetesimals that migrated from each of the regions between 0.7 and 1.5 AU from the Sun and collided with the almost-formed Earth and Venus, differed for these planets probably by not more than a factor of two. The outer layers of the Earth and Venus could accumulate the same, for these planets, material from different parts of the feeding zone of the terrestrial planets. At the final stages of the formation of the terrestrial planets, planetesimals which initially had been at a distance between 1.1 to 2.0 AU from the Sun could be incorporated by the Earth and Mars in the ratio close to the mass ratio of these planets.

The formation of the Martian embryo, the mass of which is several times smaller than that of Mars, as a result of contraction of a rarefied condensation may explain the relatively rapid growth of the bulk mass of Mars. We may also suppose that Mercury's embryo with a mass of 0.02 of the Earth's mass was the result of contraction of a condensation. The masses of this order for the embryos of Mars and Mercury, which resulted from contraction of condensations, have never been hypothesized earlier.

The portion of planetesimals that fell onto the Sun could exceed 10%, if the initial distances of planetesimals from the Sun are in ranges of 0.3−0.5 and 1.1−2.0 AU. The portion of planetesimals ejected from the feeding zone of the terrestrial planets into hyperbolic orbits did not exceed 10%. The probability of a collision with Jupiter for a planetesimal that initially was in the feeding zone of the terrestrial planets was not more than several percent of the probability of its collision with the Earth, while the probabilities of collisions of planetesimals with Saturn were on average an order of magnitude less than those with Jupiter.

The above estimates of the formation of embryos of the terrestrial planets are based on calculations within the model taking into account the gravity of the giant planets and embryos of the terrestrial planets. Accounting for the mutual gravitational influence of planetesimals may intensify the material mixing in the feeding zone of the terrestrial planets and increase the probability of collisions of planetesimals with the Sun and their ejection into hyperbolic orbits.

For the mass ratio of the embryos of the Earth and the Moon equal to 81 (i.e., the mass ratio for the Earth and the Moon), the ratio of the probabilities of infalls onto the embryos of the Earth and the Moon did not exceed 54 in the considered variants of calculations; and it was highest for the embryo masses roughly three times smaller than the present masses of these celestial bodies.

Peculiar features in the formation of the terrestrial planets can be explained even under relatively gentle decrease of the semi-major axis of Jupiter due to the ejection of planetesimals into hyperbolic orbits; and it is not necessary to consider the migration of Jupiter to the orbit of Mars and back, as in the Grand Tack model, and the sharp changes in the orbits of the giant planets falling into a resonance, as in the Nice model. In recent years, the formation of the terrestrial planets has been mainly considered within these two models.

## ACKNOWLEDGMENTS

The author is grateful to A.B. Makalkin and I.N. Ziglina for valuable comments that improved the paper.

## FUNDING

The study of accumulation of Venus, Mars, and Mercury was supported in part by Program 12 for Fundamental Researches of the Presidium of the Russian Academy of Sciences. The study of the Earth−Moon system formation was supported by the Russian Scientific Foundation, project no. 17-17-01279.

*Translated by E. Petrova*